\input epsf

\documentclass[printer]{aa501}
\usepackage[francais,english]{babel}
\usepackage{babel}

\usepackage[cp850]{inputenc} 

\usepackage[T1]{fontenc}
\usepackage[cp850]{inputenc}
\usepackage[T1]{fontenc}
\usepackage{amsmath,amssymb}
\usepackage[dvips]{graphicx}
\begin{document}


\title{On the possible existence of a self-regulating hydrodynamical
process in slowly rotating stars :}

\subtitle{ II. Lithium Plateau in Halo Stars and Primordial Abundance}

\author{Sylvie Th\'eado and Sylvie Vauclair }

\titlerunning{lithium in halo stars}

\authorrunning{Sylvie Th\'eado and Sylvie Vauclair}

\offprints{Sylvie Th\'eado}

\institute{ Laboratoire d'Astrophysique, 14 av. Ed. Belin, \\
               31400 Toulouse, France}

\date{Received 26 December 2000 / Accepted 25 May 2001}


  \abstract{
The lithium plateau observed in halo stars has long appeared as a
paradox in the 
general context of the lithium abundance behavior in stellar outer
layers. First, 
the plateau is flat, second, the lithium abundance dispersion is
extremely small. 
This seems in contradiction with the large lithium variations observed
in younger 
stars. It is also difficult to understand theoretically : as lithium
nuclei are 
destroyed by nuclear reactions at a relatively low temperature 
($\cong~2.5$ million degrees), the occurrence of macroscopic motions in the
stellar 
outer layers easily lead to lithium depletion at the surface. On the
other hand, 
if no macroscopic motions occur in the stellar gas, lithium is subject
to 
microscopic diffusion which, in the case of halo stars, should also lead
to depletion. 
Several ideas have been proposed to account for the lithium behavior in
halo stars. 
The most promising possibilities were rotational-induced mixing, which
could reduce
 lithium in the same way for all the stars (Vauclair \cite{Vauclair88}, Pinsonneault
et al. \cite{Pinsonneault92} 
and \cite{Pinsonneault99}) and mass-loss, which could oppose the
lithium settling (Vauclair \& Charbonnel \cite{Vauclair95} and \cite{Vauclair98}). In both cases however, the parameters should
be tightly adjusted 
to prevent any dispersion in the final results. 
Vauclair (\cite{Vauclair99}) (paper I) looked for a physical process
which could occur in slowly 
rotating stars and explain why the dispersion of the lithium 
abundances in the halo stars' plateau is so small. She pointed out
that the $\displaystyle \mu $-gradient terms which appear in the
computations of the meridional circulation velocity
(e.g. Mestel \cite{Mestel53}) were 
not introduced 
in previous computations of rotationally-induced mixing. This can lead
to a 
self-regulating process which reduces the efficiency of the
meridional circulation as well as the microscopic diffusion.
Here we present numerical computations of this process and its influence
on the 
lithium abundance variations in halo stars. 
We show that in slowly rotating stars, 
under some conditions, lithium can be depleted
by a factor of up to two with a dispersion smaller than 0.1 dex in the
middle part of the lithium plateau. 
We derive a primordial 
lithium abundance of 2.5$\pm$0.1, consistent with the recent determinations of
D/H and $^4$He/H.
}
\maketitle

\section{Introduction}

Prior to the first lithium detection in halo stars by Spite \& Spite
(\cite{Spite82}), 
the primordial $^7$Li value was believed to lie around A(Li) = 3.3 
(in the $\log$~H~$=~12$ scale), which corresponds to both the meteoritic
and the maximum lithium abundance in Pop~I stars throughout the 
entire observed $\displaystyle T_{{\rm eff}}$ range (8000-4500 K). Below
this 
maximum value, the observations revealed a large scatter in the 
Li - $\displaystyle T_{{\rm eff}}$ plane, with an evidence of lithium
decrease 
with age for stars of similar $\displaystyle T_{{\rm eff}}$.

For this reason it was expected that, due to their age, halo stars
should have 
destroyed all their lithium. In this context, the first lithium
detection in 
these stars, with a value A(Li)= 2.1, only one order of magnitude
smaller than 
the maximum Pop~I value, was a surprise. The shock was still stronger
when 
the observations in different stars, with different masses and
metallicities, 
presented the same lithium value over a range of 800~K in effective
temperature 
(6250-5480 K). 

From then on, the observers switched to another extreme : the idea that
lithium 
should be depleted in these stars was completely rejected on the basis
of the 
constancy and very small dispersion of its abundances. The observed
value was 
then supposed to represent the primordial abundance, without any
modification, 
and was ''sold'' as such to cosmologists. Meanwhile, theoreticians went
on 
claiming that, from all possible computations of stellar structure, the
lithium 
abundance could not keep exactly the same value in stars during 14
billions years, 
without any alteration.

In the present paper, we argue that we can reconcile the theory with the
observations and obtain a consistent view of the lithium behavior in
stars. 

In section 2, we discuss recent improvements on the observational side
and 
show that the careful studies of the lithium abundances in the `` Spite
plateau'', 
obtained from various authors, lead to the conclusion that the
dispersion is 
smaller than 0.1 dex. Meanwhile, the detection of $^6$Li, if confirmed, 
puts strong constraints on the possible lithium destruction by nuclear
reactions in these stars (otherwise no $^6$Li should be left). 

A theoretical discussion about previous models for lithium abundance
variations in population II stars is given in section 3.

In section 4, we show how the introduction of the effect of 
$\displaystyle \mu $-gradients in the computations of rotation-induced
mixing, as discussed by Vauclair (\cite{Vauclair99}) (paper I)
modifies the previously
published results. We argue that a self-regulating process may take 
place, which may explain most of the lithium observations in halo stars.

Section 5 is devoted to the results obtained from numerical computations
of this process and its influence on the lithium abundance variations in
different stars with various masses, rotational velocities and
metallicities. 

We give our conclusions in section 6, including a discussion about the
assumptions introduced in our computations. In the framework of our
model, we obtain a primordial lithium abundance of 2.5$\pm$0.1.

\section{The lithium plateau : observational constraints}

Since the first observations of the
lithium plateau by Spite \& Spite (\cite{Spite82}), many abundance determinations
have confirmed the constancy and very small dispersion of the lithium
value 
in most halo stars 
with effective temperatures larger than 5500~K.  
This result lead the observers to the conclusion that the observed
lithium abundance 
was exactly the pristine value, although theoretical computations
predicted
some lithium depletion in these stars. We first give a short summary
of the important observational constraints and then we will discuss the
theoretical models proposed in the literature.

\subsection{Trends with temperature and metallicity}

The observed lithium plateau is flat, at least to a first approximation.
Some doubts still
appear, however, about possible small variations with effective
temperature 
and/or metallicity.

Thorburn (\cite{Thorburn94}) found that the lithium
plateau presented a positive slope with $\displaystyle T_{{\rm eff}}$
which, she argued,
would not exist if the lithium was
primordial. She also
found a general trend of the lithium abundance with metallicity which
could explain part of the scatter that she observed on the ``Spite
plateau''. She suggested that the slight increase of the lithium abundance
with [Fe/H] could be due to a small production of lithium in the early
life of
the galaxy. 

While Norris et al. (\cite{Norris94}) and Ryan et al. (\cite{Ryan96}) confirmed these results,
Molaro et al. (\cite{Molaro95})
argued that the slope of the plateau disappears when a subset of stars
with temperature based on Balmer lines profiles is adopted. They
found no correlation between lithium abundances and metallicity for
[Fe/H] $<-1.4$. 

Later on, Bonifacio \& Molaro (\cite{Bonifacio97}) studied a sample of 41 plateau
stars, to investigate in more detail
the distribution of lithium abundances with
[Fe/H] and $\displaystyle T_{{\rm eff}}$. 
They used a new effective temperature scale based on the new
accurate infrared flux method (IRFM, Blackwell et al. \cite{Blackwell90}) applied to a
large sample of stars by Alonso et al. (\cite{Alonso96}).
They found only
a tiny trend with $\displaystyle T_{{\rm eff}}$ and no trend with
[Fe/H]. 
Using mean lithium values corrected for standard depletion
and NLTE effects, they give for the primordial abundance :
$A(Li_0) = 2.238 \pm 0.012$.

Meanwhile Ryan et al. (\cite{Ryan99}) studied
23 very metal-poor stars chosen to enable a precise measurement of the
dispersion in the ``Spite plateau'' : they found no evidence for a trend 
with $\displaystyle T_{{\rm eff}}$ but they do recover
a strong dependence of the lithium abundance with metallicity,
which they attribute to the chemical evolution in the early galaxy.
The value they derive for the primordial abundance,
taking into account galactic chemical evolution but
no depletion, lies below the mean plateau value :
2.0 dex only, with a systematic error
of 0.1 dex. 

While the question of slight lithium slopes with effective temperature
and/or metallicity is not yet completely settled, everyone agrees on the
fact that some stars otherwise indistinguishable from normal plateau
stars 
show large lithium deficiencies, well below the plateau itself
(Hobbs et al. \cite{Hobbs91}, Thorburn \cite{Thorburn94}, Norris et
al. \cite{Norris97}).
Although these stars are certainly lithium depleted, so that their
observed 
abundances are quite different from the pristine one, they must be
accounted for
in the theoretical scenarios.
Norris et al. (\cite{Norris97}) measured the abundances of 14 elements (Fe, Mg, Al,
Si, Ca, Sc, Ti, Cr, Mn, Co, Ni, Sr, Y, Ba) for the metal poor 
stars G66-30, G139-8 and G186-26, which are highly lithium deficient
and lie near the main sequence cut-off. They found no
abundance anomalies for the other elements that one might associate
with the lithium
deficiencies. This result is in favor of a nuclear lithium destruction
and negligible element settling in these stars.

\subsection{Lithium abundance dispersion in the plateau}

Since 1982, many authors have discussed the possible
existence of an intrinsic dispersion in the ``Spite plateau''. 
Deliyannis et al. (\cite{Deliyannis93}), 
supported  by Thorburn (\cite{Thorburn94}), argued that the spread in
measured plateau stars abundances exceeded that expected from
observational data. They gave values of about 20\% (Deliyannis et al. \cite{Deliyannis93})
to 25\% (Thorburn \cite{Thorburn94}).
Molaro et al. (\cite{Molaro95}), Spite et al.(\cite{Spite96}) and
Bonifacio \& Molaro (\cite{Bonifacio97}) have questioned
whether some of the error estimates in these earlier
works were realistic and have suggested that the dispersion is smaller.
According to Molaro et al. (\cite{Molaro95}), all the
observations are consistent with the same pristine lithium abundance and
the errors estimated for individual stars account for the observed
dispersion. Spite et al. (\cite{Spite96}) argued that the
scatter is fully explained by the temperature equivalent width errors
and that the intrinsic scatter, if real, is small.
Bonifacio \& Molaro (\cite{Bonifacio97}) revised the lithium abundances in a selected sample
of halo stars using the infrared flux method (IRFM) temperatures by
Alonso et al. (\cite{Alonso96}). They found no evidence for intrinsic
dispersion.

In the sample of Ryan et al. (\cite{Ryan99}), 21 stars out of 23
have abundances consistent with an observed spread of a mere 0.031 dex
with reference to the metallicity trend that they claim to observe. 
Because the formal errors are
0.033 dex, they conclude that the intrinsic spread of the Li
abundances at a given metallicity is lower than 0.02
dex and consistent with zero at the very metal poor halo turnoff. This
maximum dispersion is much lower than those obtained by previous 
studies (Spite et
al. $\simeq$ 0.06-0.08 dex, Bonifacio and Molaro 0.07 dex) who did not
introduce any systematic variation with metallicity.

The situation is different in globular clusters where the lithium
abundances in stars around the turn-off
seem to show a dispersion larger than that observed
for field stars.
Deliyannis et al. (\cite{Deliyannis95}) and Boesgaard et al. (\cite{Boesgaard98}) observed the
stars near the turnoff of the old metal poor globular cluster
M92.  They reveal a spread in abundances
of a factor of 2-3 for a small sample of stars. Moreover some of them show a
high lithium abundance compared to the halo
field plateau. 

These variations inside globular clusters may be related to
local pollution and other effects which will not be discussed here.
In any case the very small lithium dispersion in halo field 
stars 
(probably less than 10\% from the most recent studies),
acts as a strong constraint
on the theoretical models. 

As we will see below, these results argue against strong lithium
depletion in plateau stars. Recent observations of $^6$Li
confirm this idea.
$^6$Li has indeed been observed in the
atmosphere of the halo stars HD 84937 by Smith et al. (\cite{Smith93}),
Hobbs \& Thorburn (\cite{Hobbs94}), Hobbs \& Thorburn (\cite{Hobbs97}) and
BD +$26^o$ 3578 by Smith et al. (\cite{Smith98}).
As this light lithium isotope is more easily destroyed than $^7$Li, 
these observations, if confirmed, seem to prove that no strong 
lithium destruction by nuclear reactions occured in these stars
during main sequence or pre-main sequence. Small destruction rates may
however be allowed : we will see that a $^7$Li depletion by a factor of
two leads to a $^6$Li decrease by less than a factor of five.

\section{The lithium plateau : theoretical discussion}

The expected lithium depletion in halo stars is primarily due 
to the diffusion processes 
which take place in the
radiative regions inside the stars, below the outer convective zones.
Due to pressure and temperature gradients, lithium settles down, as well
as 
helium and other heavier elements.
Mixing induced by rotation, internal waves, or mass loss related
motions may slow down the settling process, but then it brings up to the
convective zone matter in which lithium has been destroyed by nuclear
reactions.

These lithium depleting processes have been extensively studied in the
literature. Michaud et al. (\cite{Michaud84}) first computed the lithium abundance
variations in halo stars including element separation. Their models 
predicted a downward curvature in the $^7$Li  isochrones for increasing
effective temperature, in contradiction with the observations. This
result, 
simply due to the decrease of the convective depth for hotter stars, 
has been confirmed many times in the literature (Deliyannis et
al. \cite{Deliyannis90}, Deliyannis \& Demarque 
\cite{Deliyannis91}, Proffitt \& Michaud \cite{Proffitt91}, Chaboyer
\& Demarque \cite{Chaboyer94}, Vauclair \& Charbonnel
\cite{Vauclair95}). 

Models including rotation-induced mixing were proposed to prevent the 
effect of element separation. As such a mixing leads to lithium 
destruction by nuclear reactions, it was suggested that this destruction
could be identical for all the stars, so that the lithium plateau could
be 
preserved (Vauclair \cite{Vauclair88}, Pinsonneault et
al. \cite{Pinsonneault92}, \cite{Pinsonneault99},
Chaboyer \& Demarque \cite{Chaboyer94}). However, 
if a constant lithium abundance can be obtained in this case, it becomes
extremely difficult to account for such a small dispersion as observed.

Stellar models which include instabilities related to angular momentum
and 
chemical species transport require additional input physics beyond 
standard models. The important new ingredients include :

1- a distribution of initial angular momenta

2- a description for angular momentum losses

3- a prescription for the internal transport of angular momentum
and for the associated mixing processes in the stellar radiative regions.

Using various initial angular momenta, Pinsonneault et
al. (\cite{Pinsonneault92}) obtained lithium depletion factors between 5
and 10 at fixed $\displaystyle T_{{\rm eff}}$.
The lithium abundances in the plateau slightly decreased for larger 
$\displaystyle T_{{\rm eff}}$ in contradiction with the observations
(this trend was already predicted by Vauclair (\cite{Vauclair88})). This effect
however is quite attenuated for older ages (14 billion years) as the
more 
lithium depleted stars at the hot end
of the plateau move into the giant branch.

Chaboyer \& Demarque (\cite{Chaboyer94}) could also reproduce a lithium plateau
with a large lithium depletion 
in a similar framework, using
different angular momentum loss and rotational mixing laws. They
included
meridional circulation, dynamical shear, secular shear, Solberg-Hoiland
and
Goldreich-Schubert-Fricke instabilities. 
Their results also showed the same trend of abundance decrease with
increasing $\displaystyle T_{{\rm eff}}$.

Using a distribution of initial angular momenta as inferred from stellar
rotation data in young open clusters, Pinsonneault et al. (\cite{Pinsonneault99}) obtained
a well defined, nearly flat lithium plateau, without the downward trend 
at the hot end. A modest scatter does remain however, which increases with
the
average lithium depletion : this constraint excludes a lithium
depletion larger than a factor of 3.

Considering the difficulty for theoretical computations, including 
element settling and rotation induced mixing, to account for
both the flatness of the lithium plateau and the small dispersion,
Vauclair \& Charbonnel (\cite{Vauclair95}) and Vauclair \& Charbonnel
(\cite{Vauclair98}) suggested taking into account a stellar 
wind. In some cases, a wind could prevent the element
settling without bringing up to the convective zone lithium depleted
matter. Computations were done for various mass loss rates, between
1 and 1000 times the solar wind. Rates larger than 10 times the solar
wind
were needed to have any effect on the lithium depletion, and the best 
results were obtained for
rates of about $10^{-12} M_{\odot}.yr^{-1}$. In this case the lithium
plateau
was nicely reproduced. For larger rates, nuclearly depleted matter began
to migrate up into the outer layers, rapidly leading to large lithium
destruction, in contradiction with the observations.
Although appealing, this model suffers from the fact that it needs
winds larger than the solar wind, and that various rates lead 
to a non-negligible scatter in the results.

The fact that the dispersion in the lithium plateau
is extremely small compared to the observational 
errors is the strongest constraint for the theoretical models.
Several assumptions have lead to models which could account for the
lithium abundance constancy, but the small observed scatter could only
be
reproduced using some 
``ad hoc'' hypothesis on the initial parameters. 

In all these computations, however, the effect of the 
$\displaystyle \mu $-gradients induced by element separation 
was not taken into account.
In the present paper, we claim that when it is introduced in 
the computations of rotation-induced mixing, this process can 
explain the features observed in halo stars, including the very small
dispersion.

\section{Diffusion and mixing in the presence of
$\displaystyle \mu $-gradients}

The meridional
circulation velocity in stars, in the presence of 
$\displaystyle \mu $-gradients, is the sum of two terms, one due to
the classical thermal imbalance ($\displaystyle \Omega$-currents) and
the other one due to the induced horizontal
$\displaystyle \mu $-gradients ($\displaystyle \mu $-induced
currents, or $\displaystyle \mu $-currents in short). In the most
general cases, $\displaystyle \mu $-currents are opposite to
$\displaystyle \Omega$-currents (Mestel \cite{Mestel53}, Zahn
\cite{Zahn92}, Maeder \& Zahn \cite{Maeder98}).
When element settling occurs 
below the stellar outer convective zone in cool stars, a small helium 
gradient builds, even in the presence of circulation.

Then a new
process must take place, which had not been taken into account in
previous computations of diffusion-induced $\displaystyle \mu
$-gradients. Mestel \& Moss (\cite{Mestel86}) gave a lengthy discussion of this
effect for nuclearly-induced $\displaystyle \mu $-gradients. Chaboyer
et al. (\cite{Chaboyer95}) simulated a $\displaystyle \mu $-gradient effect in
their diffusion computations in the form of a reducing factor
f$_{\mu}$ which was taken as a parameter. 
Vauclair
(\cite{Vauclair99}) (paper~I) showed that 
the resulting $\displaystyle \mu $-gradients are rapidly
large enough to create $\displaystyle \mu $-currents of the same order
as 
$\displaystyle \Omega$-currents. Then a self-regulating process 
may take place, in which both the circulation and the settling
are strongly reduced. 

Here we give precise computations of the $\displaystyle \mu $-currents
with the following simplifying assumptions : differential rotation
is supposed negligible inside the stars (as it is presently in the
sun) and the stellar rotation velocity is taken as constant along the
stellar lifetime.
These assumptions neglect the possibility of rotational breaking in
early stages of stellar evolution. This is discussed in the conclusion.
  
\subsection{Computations of $\displaystyle \Omega $ and
$\displaystyle \mu $-currents}

In the present paper, we have computed these currents
in halo stars for different metallicities and rotation velocities
and we have done a complete treatment of the resulting diffusion 
of the chemical species, including mixing and settling.

As in paper~I, we have neglected the deviations from a perfect gas law
as well as the energy production terms, which are completely
negligible in the regions of the star where the process takes
place.

We also assumed a negligible differential rotation, as observed
inside the Sun from helioseismic studies.  
The corresponding condition on $\displaystyle \Omega$ is :
\begin{equation}
\vert {\partial \ln \Omega \over \partial \ln r }\vert 
< {\Omega ^{2}r^{3}\over GM}
\end{equation}

The vertical meridional circulation velocity may be written (paper I) :
\begin{equation}
u_{r} =
{{\bf \nabla} _{{\rm ad}}
 \over{\bf \nabla}_{{\rm ad}} - {\bf \nabla} + \nabla_{\mu }} \
{\varepsilon _{\Omega} \over g}
\end{equation}
where $\displaystyle g$ represents the local gravity, 
$\displaystyle {\bf \nabla}_{{\rm ad}}$ and 
$\displaystyle {\bf \nabla}$ the usual adiabatic and real ratios
$\displaystyle \left( {d \ln T\over d \ln P }\right)$
and $\displaystyle \nabla_{\mu }$ the mean molecular weight
contribution 
$\displaystyle \left( {d \ln \mu  \over d \ln P}
\right)$.

The expression of $\displaystyle \varepsilon _{\Omega}$ is 
obtained as a function of the $\displaystyle \Omega $ and
$\displaystyle \mu $-currents :
\begin{equation}
\varepsilon _{\Omega} =
\left( {L \over M}\right)
\left( E_{\Omega} + E_{\mu }\right) P_{2}
(\cos \theta)
\end{equation}
with:
\begin{eqnarray}{}
E_{\Omega} & = & {8 \over 3}
\left({\Omega^{2}r^{3} \over GM }\right)
\left( 1 - {\Omega^{2} \over 2\pi G\overline \rho  }
\right) \\
E_{\mu } & = &  {\rho _{m} \over \overline \rho  } 
\left\{
{ r \over 3 } \
{d \over dr }
\left[
\left(H_{T}
{d \Lambda \over dr}\right)
- (\chi_{\mu } + \chi_{T} + 1) \Lambda \right]
- {2 H_{T} \Lambda \over r } \right\}
\end{eqnarray}

Here $\displaystyle \overline \rho $ represents
the density average on the level surface 
$\displaystyle (\simeq \rho )$ while
$\displaystyle \rho _{m}$ is the mean density inside
the sphere of radius $\displaystyle r$; 
$\displaystyle H_{T}$ is the 
temperature scale height;
$\displaystyle \Lambda$  represents the 
horizontal $\displaystyle \mu $ fluctuations
$\displaystyle {\tilde{ \mu}\over \overline \mu } $;
$\displaystyle \chi _{\mu }$ and
$\displaystyle \chi _{T}$ represent the
derivatives :
\begin{equation}
\chi_{\mu } =
\left(
{\partial \ln \chi \over \partial \ln \mu  }\right)_{P,T}
\quad  ; \quad 
\chi_{T} =
\left( {\partial \ln \chi \over \partial \ln T }\right)_{P, \mu }
\end{equation}

While the term with the second derivative of $\displaystyle \Lambda$ 
was neglected in paper~I, these expressions are extensively 
computed in the present numerical computations.

Writing $u_{r}$ as :
\begin{equation}
u_{r} =
U_{r} \ P_{2} \ (\cos \theta)
\end{equation}
the horizontal meridional velocity is given by :
\begin{equation}
u_{\theta} =
- {1 \over  2 \rho r} \
{d \over dr}\
(\rho r^{2} U_{r})
\sin \theta \cos \theta
\end{equation}

In the following, we also use the notations $U_{\Omega}$ and 
$U_{\mu}$ defined as :
\begin{equation}
U_{r} = U_{\Omega} + U_{\mu} 
\end{equation}

where $U_{\Omega}$ represents the part of $U_{r}$
which includes $E_{\Omega}$, and $U_{\mu}$ the part
including $E_{\mu}$, namely : 

\begin{equation}
U_{\Omega} =
{{\bf \nabla} _{{\rm ad}}
 \over{\bf \nabla}_{{\rm ad}} - {\bf \nabla} + \nabla_{\mu }} \
{E_{\Omega} \over g} \frac{L}{M}
\end{equation}
idem for $U_{\mu}$.

The expressions and orders of magnitude of the
horizontal $\displaystyle \mu $-gradients 
$\displaystyle \Lambda$ were extensively
discussed in paper~I. 
In all cases, on a given
level surface $\displaystyle \Lambda$ is
proportional to the vertical $\displaystyle \mu $-gradient.
A general expression can be given in the form :
\begin{equation}
\Lambda =
{\tilde \mu  \over \mu  } =
\alpha_{\Lambda} \cdot r \cdot \nabla \ln \mu 
\end{equation}

where $\alpha_{\Lambda}$ is a constant, which
is related to the horizontal diffusion coefficient in case
of shear flow instabilities by (Zahn 1993, paper~I)~:
\begin{equation}
\alpha_{\Lambda} = {1 \over 6\alpha _{h}} \quad \quad  {\rm where} \quad \quad 
\alpha _{h} = {D_{h} \over (U_{r} \cdot r)}
\end{equation}

\subsection{Element diffusion and abundance
variations}

As discussed many times in the literature, stellar regions
in which no macroscopic motions take place are subject to
element settling induced by pressure and temperature gradients
and radiative acceleration ( Michaud \cite{Michaud70}, Michaud et
al. \cite{Michaud76} (MCV2), Vauclair \& Vauclair
\cite{Vauclair82}, Vauclair \& Charbonnel
\cite{Vauclair95}, \cite{Vauclair98}).
When macroscopic motions occur, they slow down the settling but
do not prevent it completely until an equilibrium concentration
gradient is reached (Vauclair et al. \cite{V2M78} (V2M),
\cite{V2SM78} (V2SM),
Richard et al. \cite{Richard96}). 

In paper~I we showed that in halo stars this equilibrium gradient  
is never reached before the $\displaystyle \mu $-currents become
of the same order of magnitude as the $\displaystyle \Omega $-currents.
Thus the abundance variations induced by 
settling and rotation-induced mixing must always be 
computed simultaneously.

Computations of element settling have been described many times. Here
we use the same prescriptions as in Richard et al. (\cite{Richard96}) with collision
cross sections and microscopic diffusion coefficients as given by 
Paquette et al. (\cite{Paquette86}). Radiative accelerations are neglected, as usual in
these cool stars (precise computations in the Sun by 
Turcotte et al. (\cite{Turcotte98}) show
that they are indeed negligible for our purpose).

Rotation-induced mixing is also computed as in Richard et al. (\cite{Richard96})
except for the introduction of the $\displaystyle \mu $-currents
in the velocity of meridional circulation. Following Zahn (\cite{Zahn92}) and 
(\cite{Zahn93}), meridional circulation is suppose to induce shear flow 
instabilities which, in a density stratified medium, lead to large
horizontal diffusivities. The combination of the circulation and
horizontal mixing may be treated as a vertical effective diffusion
process, described with an effective diffusion coefficient :
\begin{equation}
D_{{\rm eff}} \simeq {1 \over C_{h} }
U_{r} \cdot r \quad \quad  {\rm with} \quad \quad 
C_{h} \simeq 30 
\end{equation}

Abundance variations of helium, lithium and
heavier elements
are computed within this framework, all along evolutionary sequences.
The modifications of the internal structure induced by element diffusion
are taken into account in the computations, as in Richard et al. (\cite{Richard96}).

\subsection{Self-regulating process}

The originality of the present paper lies in the introduction
of $\displaystyle \mu $-currents in the meridional circulation.
As discussed in paper~I and verified in the complete
numerical computations (see below), $\displaystyle \mu $-currents
become of the same order as $\displaystyle \Omega $-currents for
small $\displaystyle \mu $-gradients, rapidly reached from 
helium settling below the convective zone.
In this case we claim that, under some conditions, a self-regulating
process may take place in which both the circulation and the
settling are strongly reduced.

At the beginning of the stellar evolution, 
$\displaystyle \vert E_{\mu } \vert$
is smaller than $\displaystyle \vert E_{\Omega} \vert$ ;
meridional circulation proceeds and induces transport
of chemical elements coupled with settling. The helium
concentration gradient below the convective zone increases (in
absolute value) leading to an increasing vertical
$\displaystyle \mu $-gradient . The horizontal 
$\displaystyle \mu $-gradient $\displaystyle \Lambda$ follows
proportionally until 
$\displaystyle \vert E_{\mu } \vert$
becomes of the same order as $\displaystyle \vert E_{\Omega} \vert$,
where the circulation
velocity is expected to
vanish. 

Then the whole process is modified and does not follow any longer the 
classical treatment of the Eddington-Sweet circulation. Let us call
$\displaystyle \Lambda_{crit}$ the horizontal 
$\displaystyle \mu $-gradient for which 
$\displaystyle \vert E_{\mu } \vert$ = $\displaystyle \vert E_{\Omega} \vert$.
Without microscopic diffusion, the whole circulation would become frozen,
keeping horizontal $\displaystyle \mu $-gradients of the order of
$\displaystyle \Lambda_{crit}$ in every layer.
However microscopic diffusion still proceeds below the convective
zone, leading to a slight decrease of $\displaystyle \Lambda$ in the
upper radiative layers (this is due to the fact that there is
no horizontal $\displaystyle \mu $-gradient inside the
convective zone because of rapid mixing). 

As $\displaystyle \Lambda$ becomes smaller than 
$\displaystyle \Lambda_{crit}$, 
circulation proceeds just below the convective zone, 
in the direction of $U_{\Omega}$. Doing so,
due to mass conservation, it lifts up matter from
below in an ascending flow and pushes it down in a 
descending flow. Although this description looks like 
that of normal
meridional circulation, there is a fundamental difference :
in the deeper layers, where $\displaystyle \Lambda$ was close
to $\displaystyle \Lambda_{crit}$, this extra motion creates 
an over-critical horizontal $\displaystyle \mu $-gradient, 
which leads to a local horizontal motion in the direction of
$U_{\mu}$. Then part of the matter which falls down in the 
descending flow is transferred to the ascending flow and
goes back up to the convective zone where it came from.
We can infer that, when this occurs, the global 
element depletion from the convective zone is reduced in a
significant way, leading to time scales larger than the stellar ages. 

A good treatment of this self-regulating process needs
a complete numerical simulation, which is planned for
a forthcoming paper. Here we used the following
approximation.

In the layers where the circulation time scale 
(that is, the time scale of
both the $\displaystyle \Omega $-currents and
the $\displaystyle \mu $-currents taken separately)
 is smaller than the settling 
time scale, we assumed that, as soon as 
$\displaystyle \vert E_{\mu } \vert$
becomes of the same order as 
$\displaystyle \vert E_{\Omega} \vert$ (within 10\%), the
$\displaystyle \mu $-gradient remains equal to the equilibrium one. 

On the contrary, in the layers where the circulation 
time scale is larger than the settling time scale, 
we assumed that
the readjustement would have 
no time to take place in a settling time scale and 
element
depletion proceeds unaltered.

In other words, the $\displaystyle \mu $-gradient remains
constant below the convective zone, where $U_{\Omega}$ is larger than
the settling velocity, as soon as $|E_{\mu }| \simeq
|E_{\Omega}|$. Meanwhile the overall helium and lithium abundances are still
subject to depletion due to diffusion deeper in the star, at the place
where the settling velocity becomes larger than $U_{\Omega}$ (see
below, Figure 2).
 
\section{Computational results}

\subsection{Evolution of $\mu$-currents with time in low metallicity stars}

First we present the results obtained for four low metallicity stars with
masses of
$0.75 M_{\odot}$ (hot end of the plateau), $0.70 M_{\odot}$ (middle of the plateau),$0.65 M_{\odot}$ (cool end of the plateau) and $0.60 M_{\odot}$ 
to show the behavior of the $\mu$-currents with time.
Here the rotation velocity is assumed constant, equal to 
$V_{rot}=5$ km.s$^{-1}$ and the metallicity is chosen equal to [Fe/H]=-2.0.  

Figures 1 (a, b, c, d) display the $|E_{\Omega}|$ and $|E_{\mu}|$
profiles below the convective zones at different evolutionary stages.
Each star arrives on the main sequence with nearly homogenous
composition. At that time,
$|E_{\mu}|$ is smaller than $|E_{\Omega}|$ everywhere inside the star.
Meridional circulation and element settling can take place 
in the radiative regions of the star as described above. 
As a consequence of the helium settling, the $\mu$-currents increase
below the convective zone and become rapidly of the same order as $\Omega$-currents.  

\begin{figure*}[p]
\centering
\includegraphics[width=13.4cm]{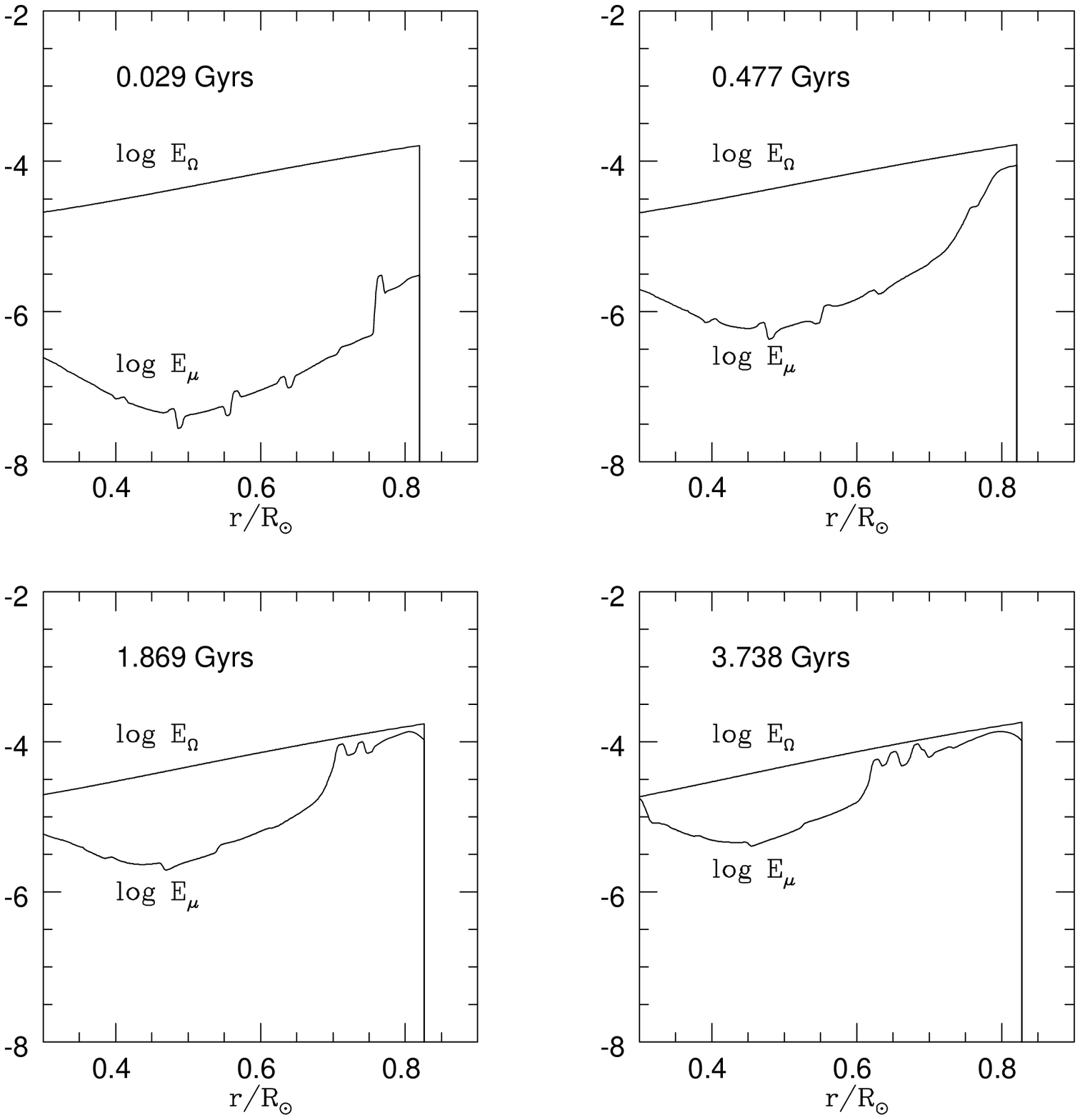}
\includegraphics[width=13.4cm]{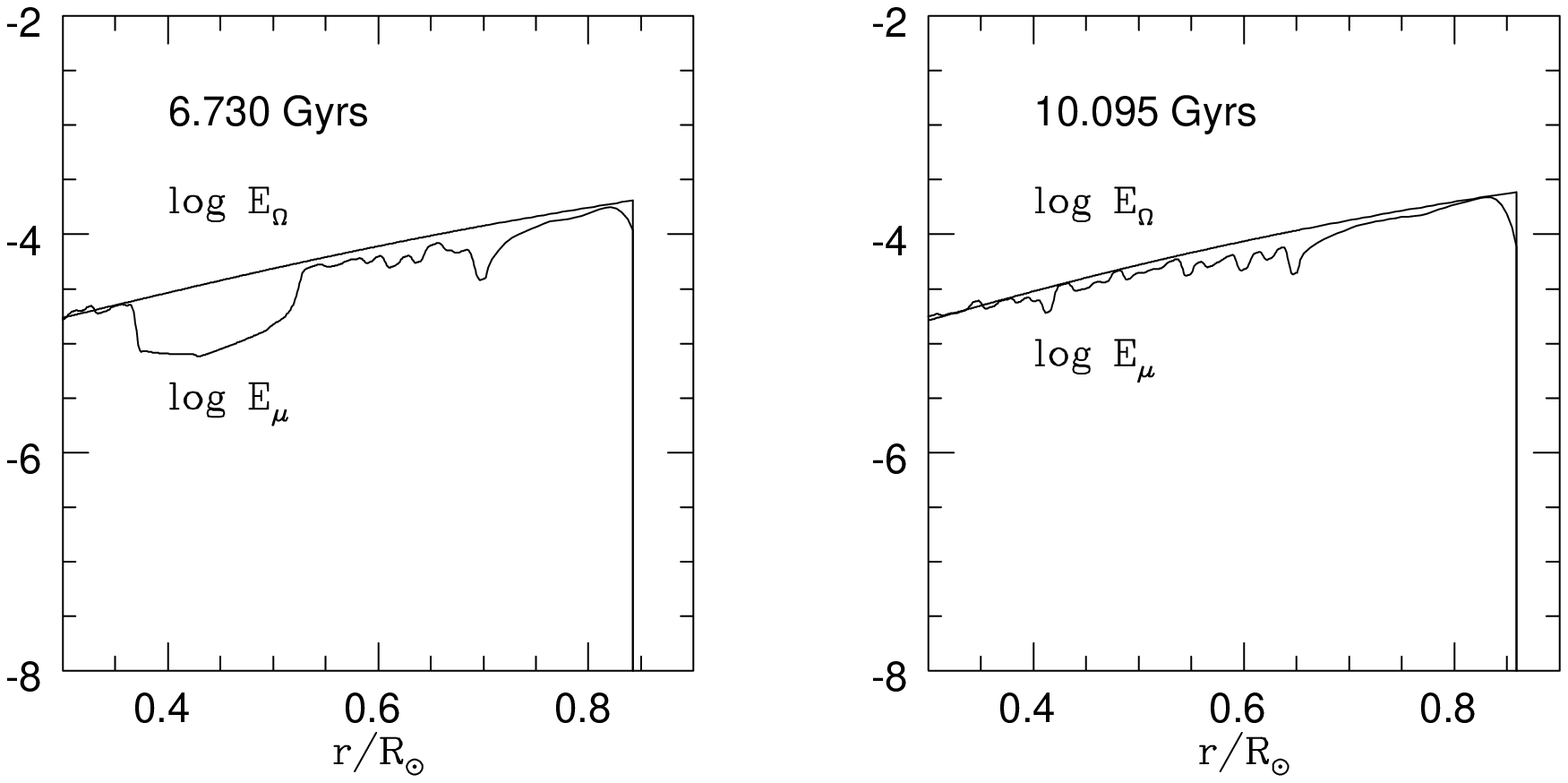}
\caption{\textbf{a.} Evolution of $\mu$-currents with time inside
the star. The graphs
show the variations with depth of both $|E_{\Omega}|$ and $|E_{\mu}|$
(defined in eq. 4 and 5) in a
$0.75 M_{\odot}$ halo star with $[Fe/H]=-2$ and $V_{rot}=5$ km.s$^{-1}$.}
\end{figure*}

\begin{figure*}[p]
\centering
\includegraphics[width=13.4cm]{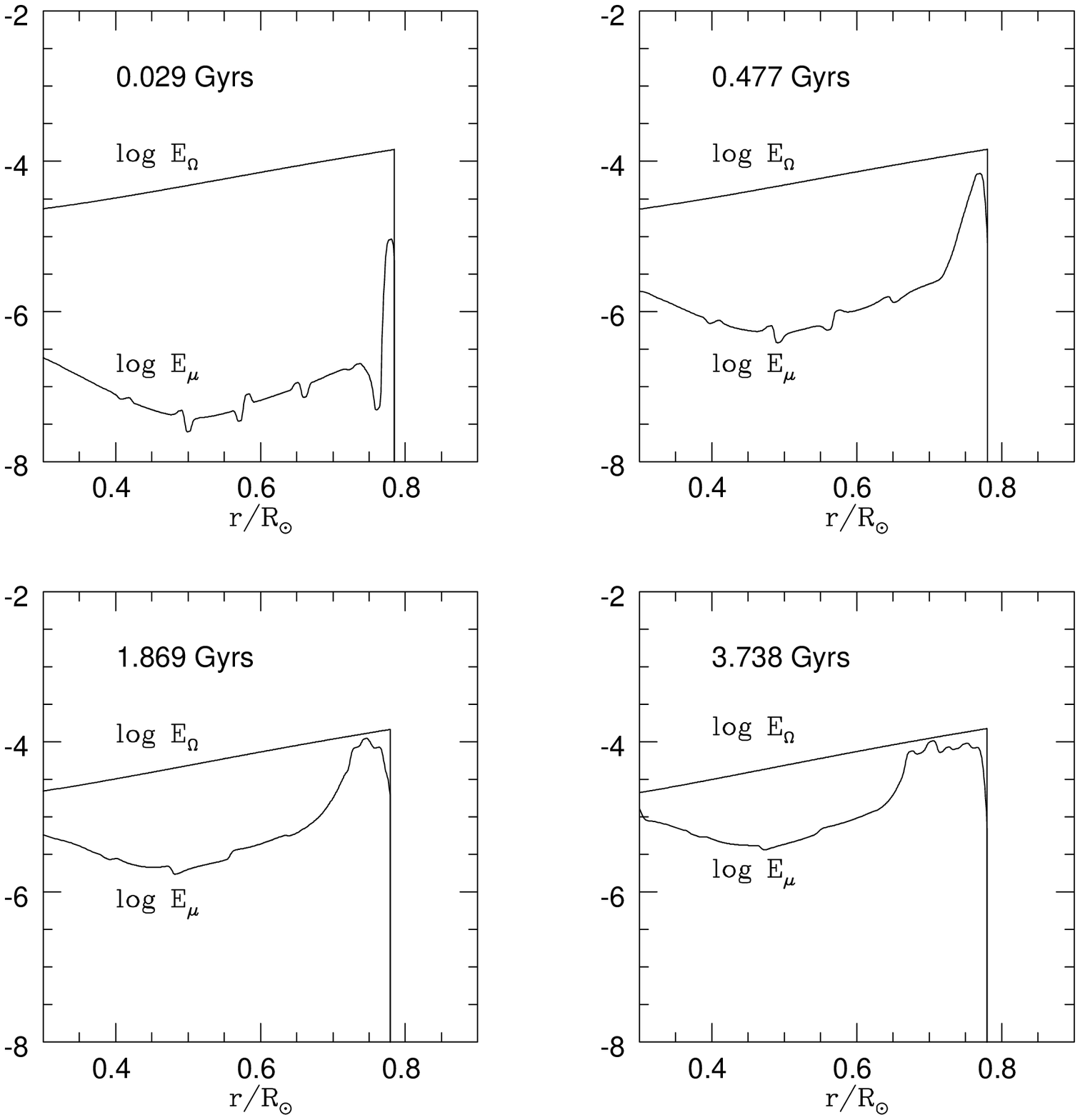}
\includegraphics[width=13.4cm]{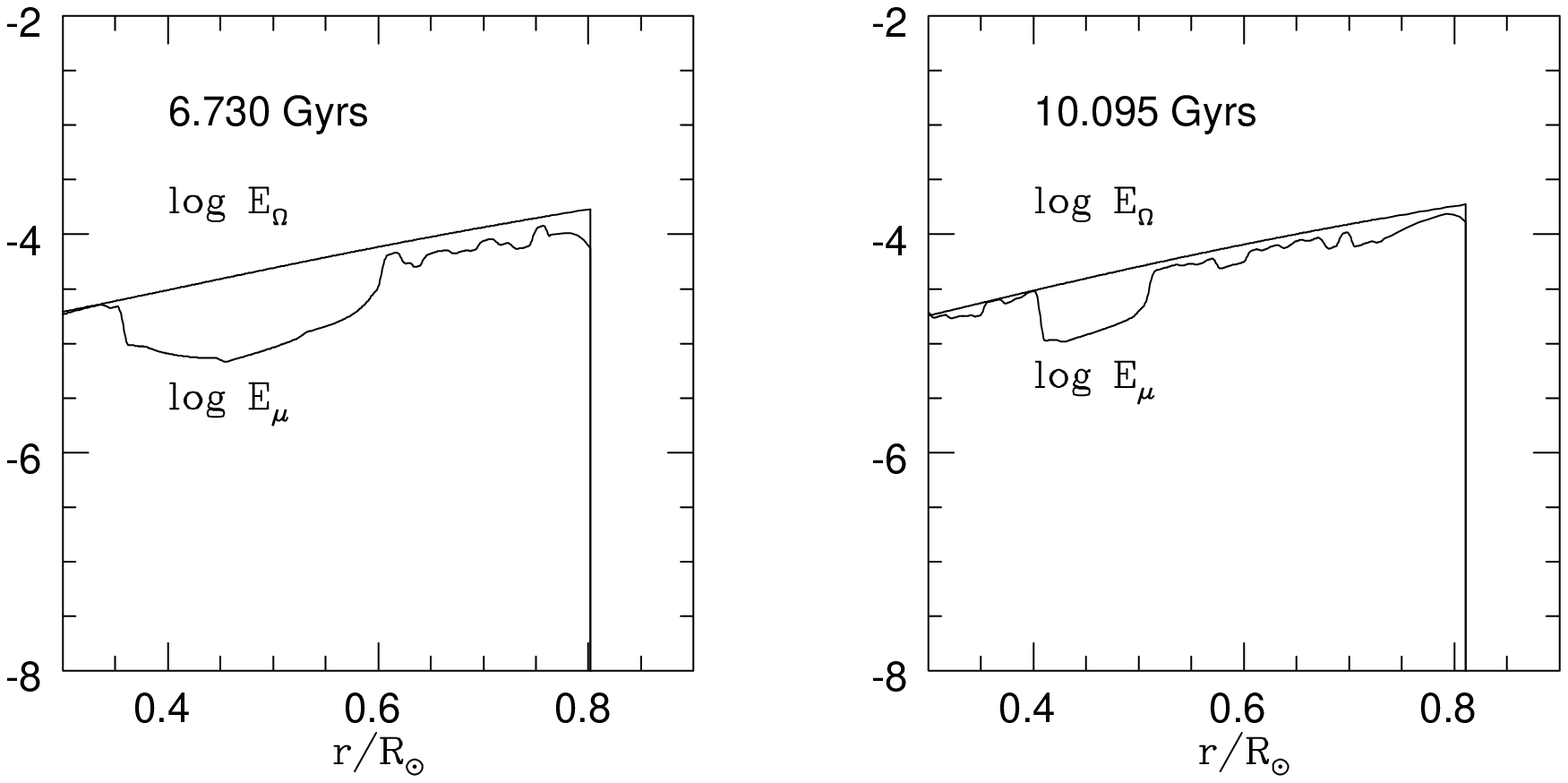}
\centerline{
\textbf{Figure 1. b.} Same figure as Figure 1.a. for a $0.70 M_{\odot}$
halo star.}
\end{figure*}

\begin{figure*}[p]
\centering
\includegraphics[width=13.4cm]{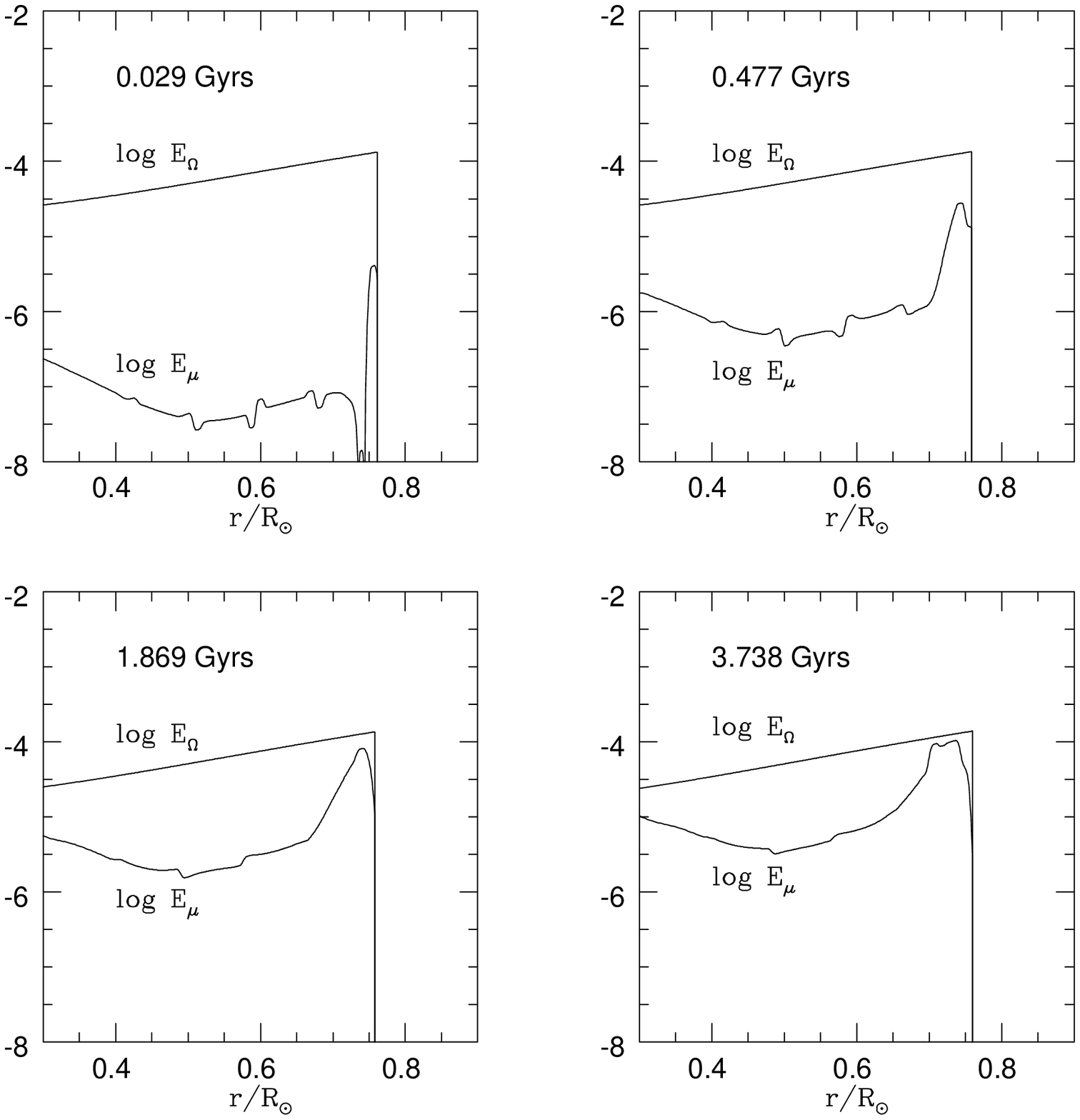}
\includegraphics[width=13.4cm]{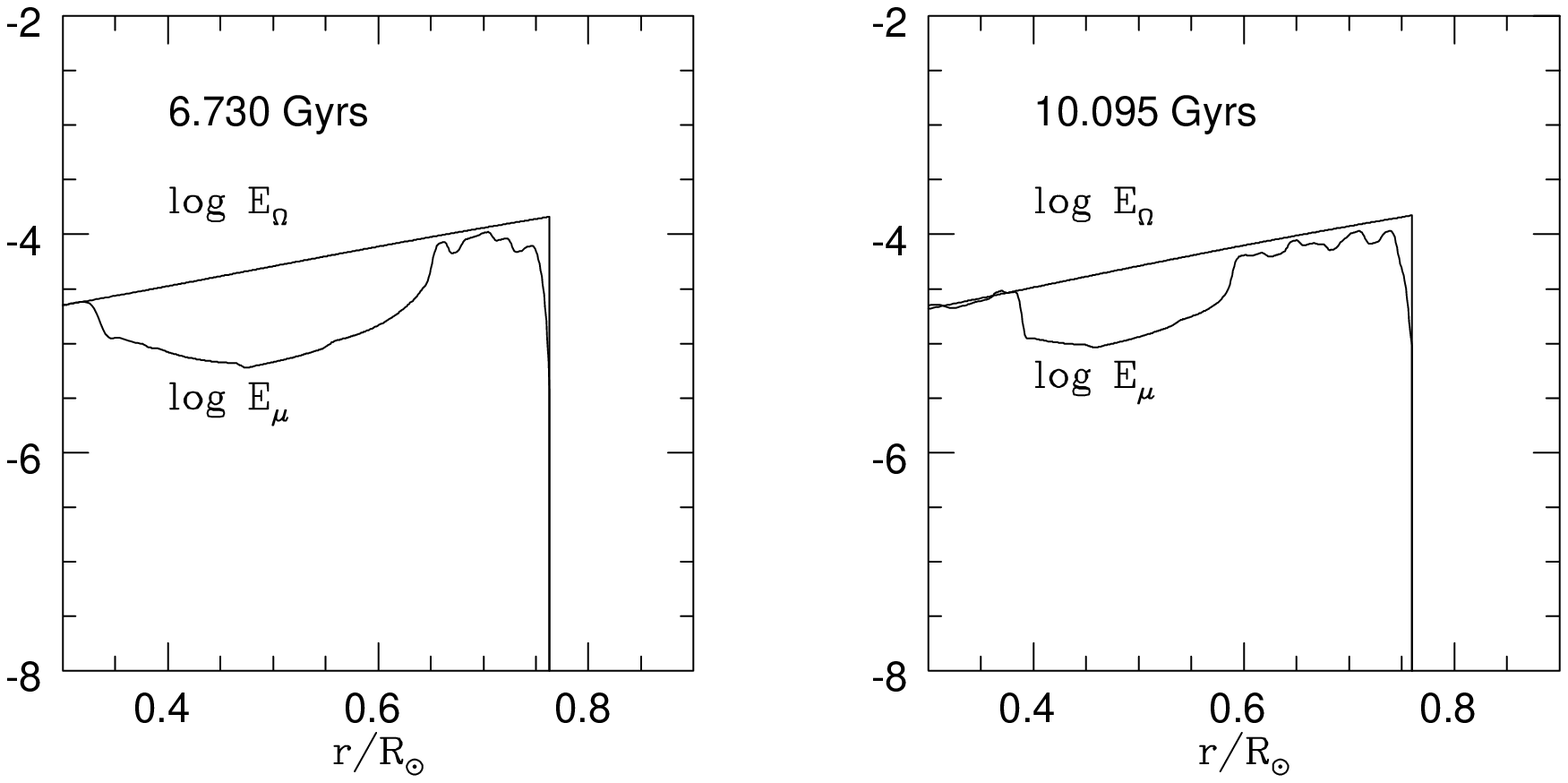}
\centerline{\textbf{Figure 1. c.} Same figure as Figure 1.a. for a $0.65 M_{\odot}$
halo star.}
\end{figure*}

\begin{figure*}[p]
\centering
\includegraphics[width=13.4cm]{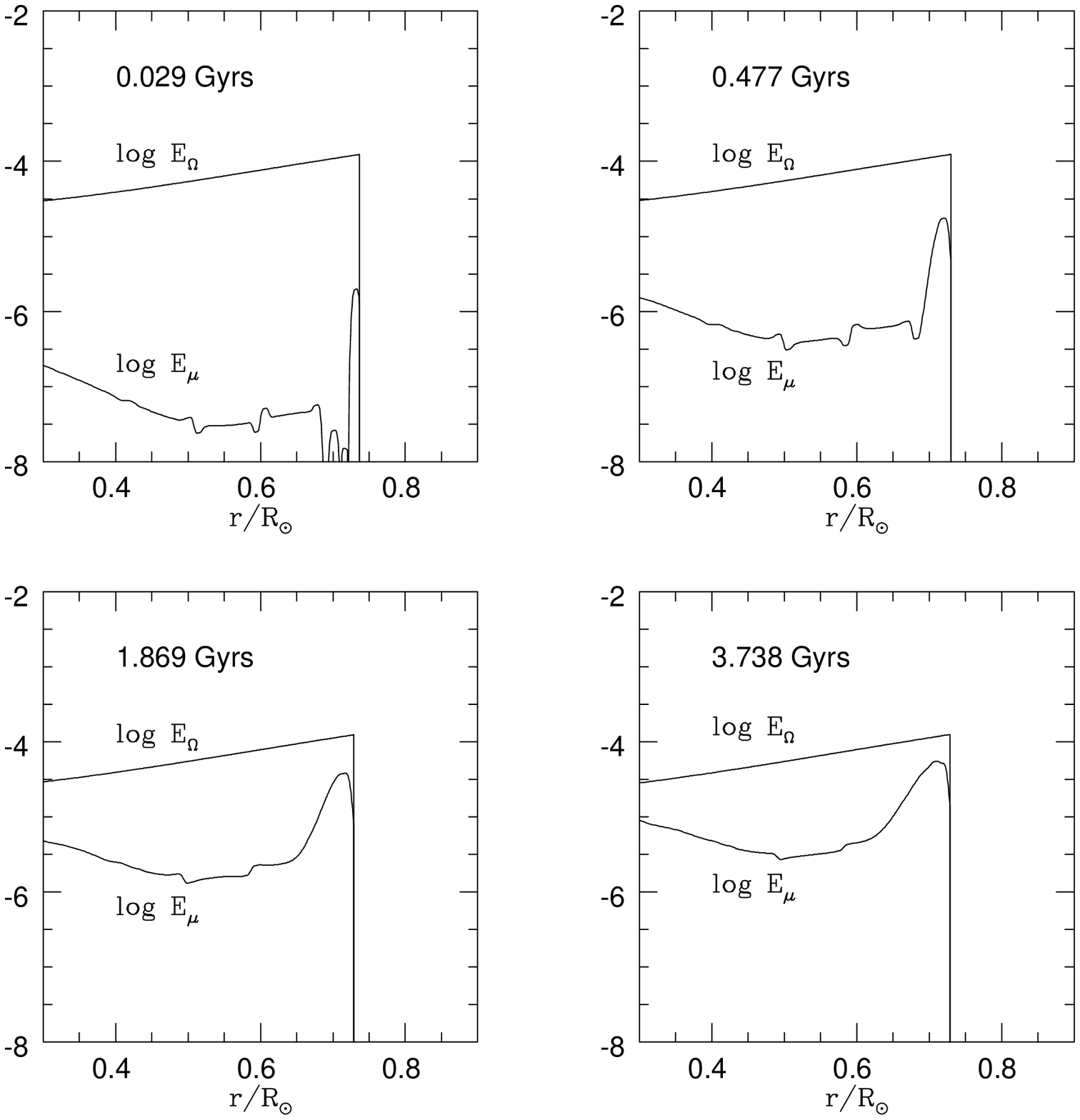}
\includegraphics[width=13.4cm]{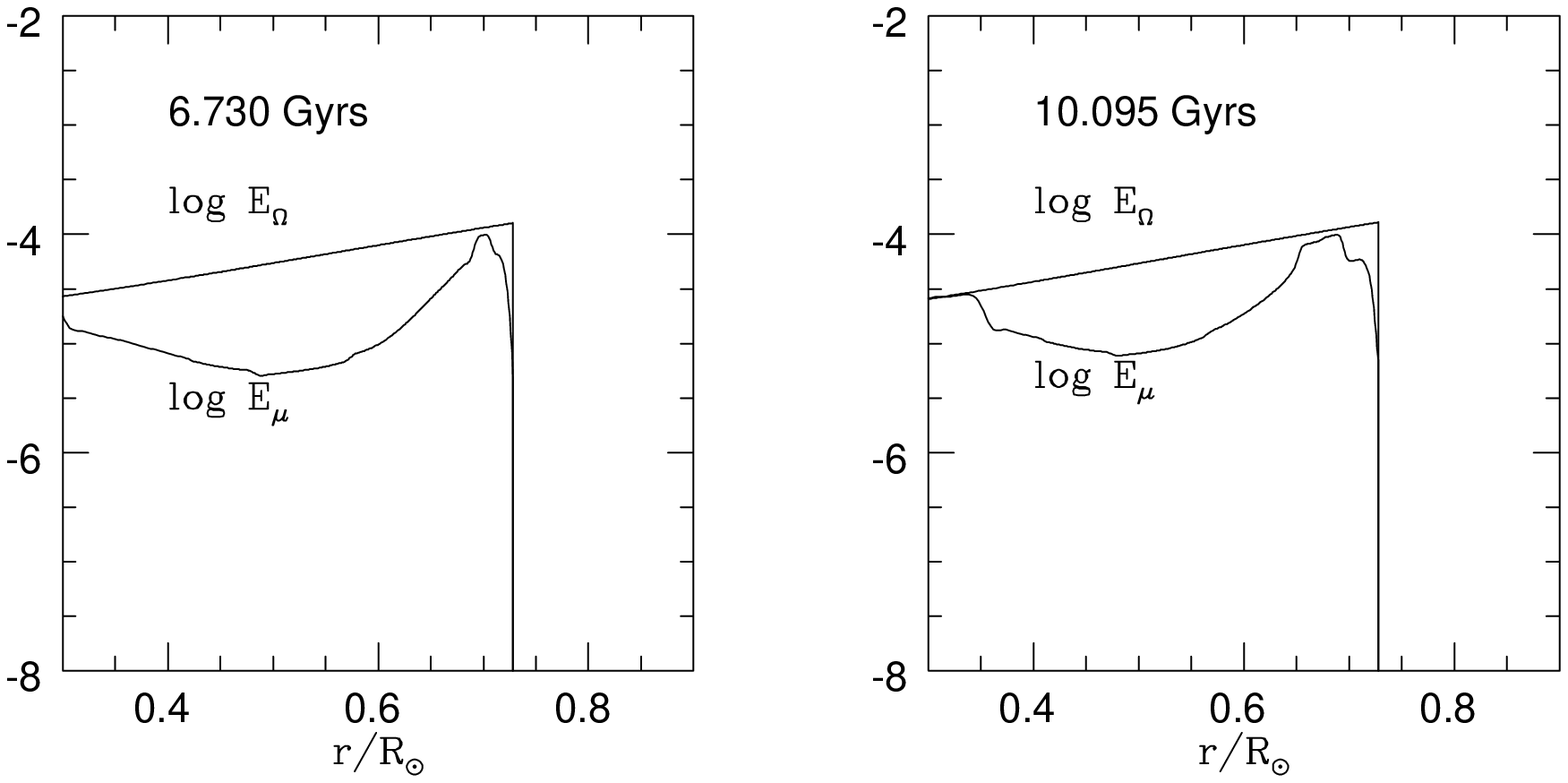}
\centerline{\textbf{Figure 1. d.} Same figure as Figure 1.a. for a $0.60 M_{\odot}$
halo star.}
\end{figure*}

\subsection{Lithium abundance variations in plateau stars}

The lithium abundance variations with time have been computed 
using the following scenario : microscopic diffusion damped by 
rotation-induced mixing occurs normally until the relative
difference between $|E_{\mu}|$ and $|E_{\Omega}|$ becomes smaller
than $\cong~10\%$. Then the self-regulating process is supposed 
to take place, in which diffusion and mixing are strongly
reduced (with time scales larger than the stellar lifetime). This
occurs down
to the place where the microscopic diffusion velocity becomes
preponderant (Figure 2, a, b, c). Below this layer, diffusion is assumed
to proceed freely.

\begin{figure*}[p]
\centering
\includegraphics[width=13.4cm]{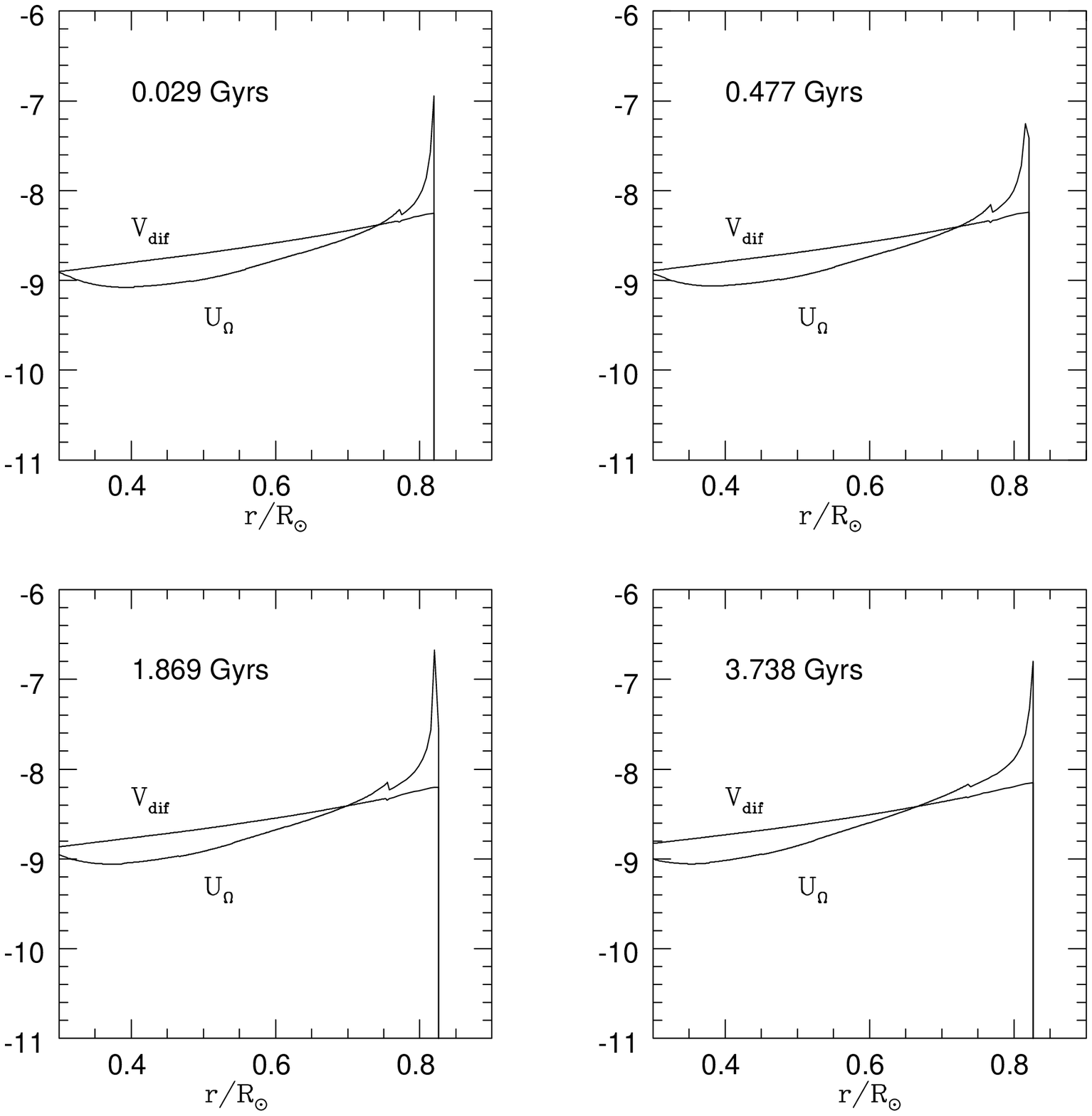}
\includegraphics[width=13.4cm]{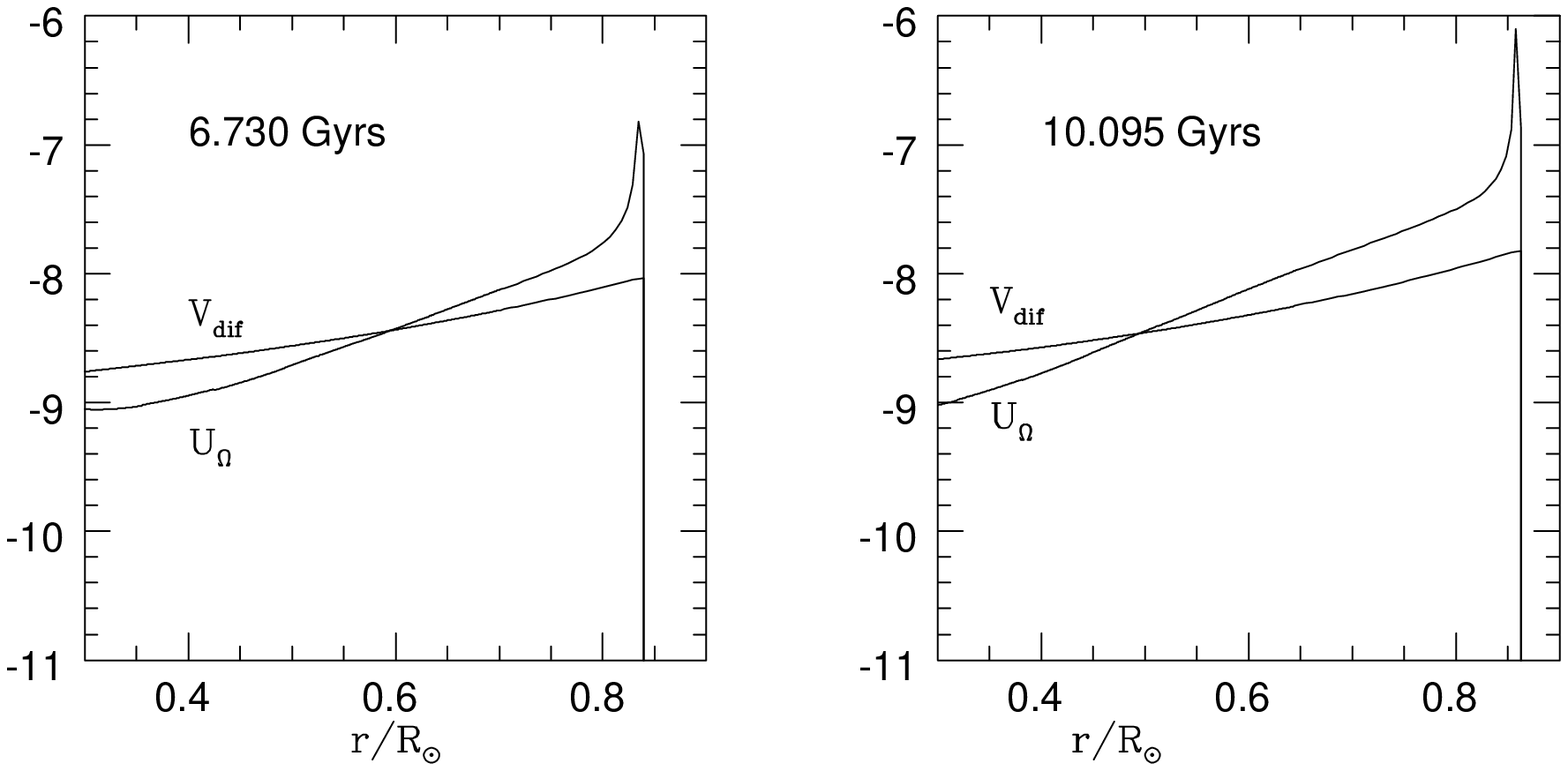}
\caption{\textbf{a.} Microscopic diffusion velocity below the convective zone
of a $0.75 M_{\odot}$ 
star compared to the classical meridional
 circulation velocity $U_{\Omega}$ (defined in eq. 10). When $V_{dif} < U_{\Omega}$,
the self-regulating process can occur while when $V_{dif} > U_{\Omega}$, 
the circulation cannot compensate for the effect of diffusion.}
\end{figure*}

\begin{figure*}[p]
\centering
\includegraphics[width=13.4cm]{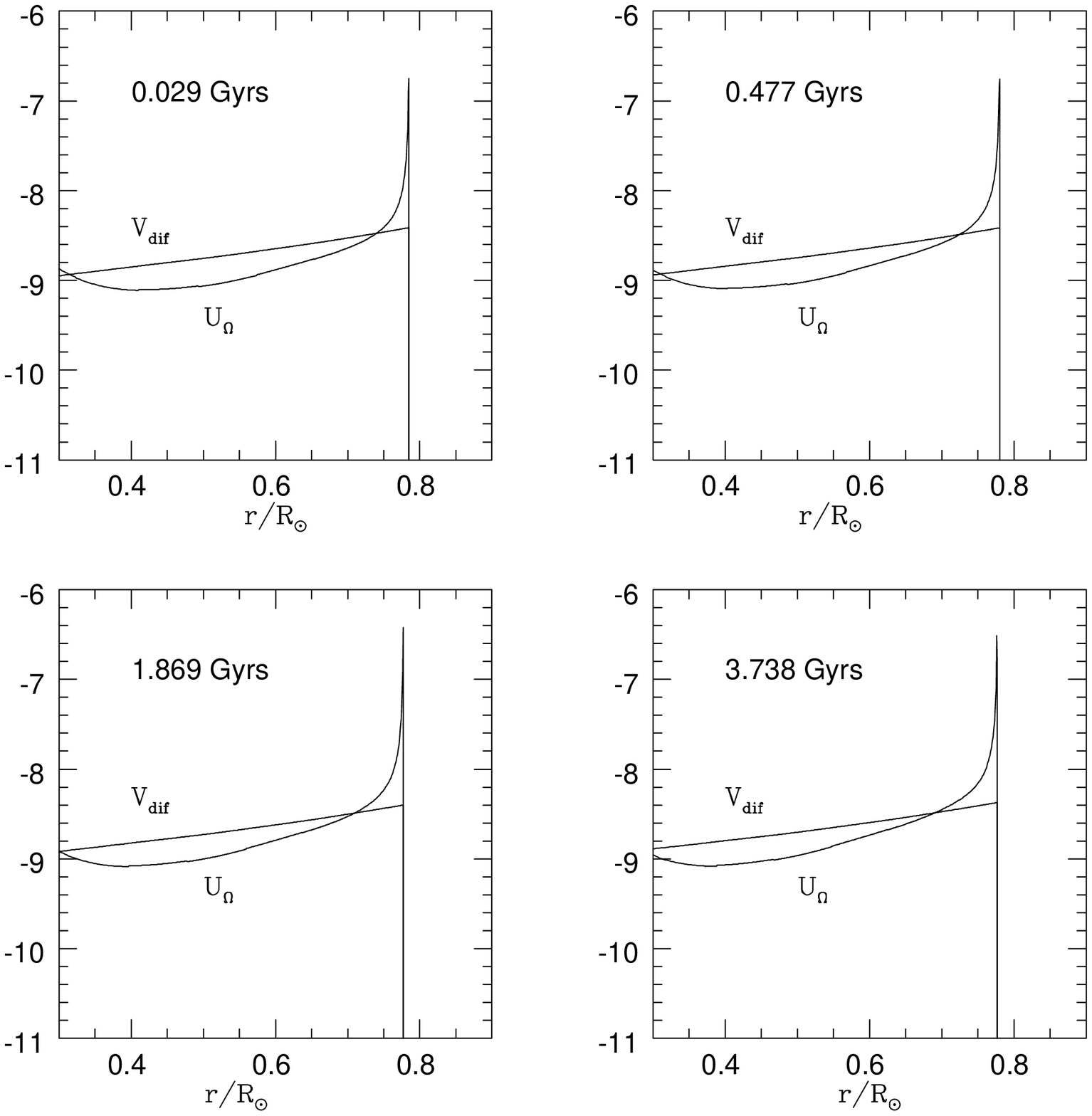}
\includegraphics[width=13.4cm]{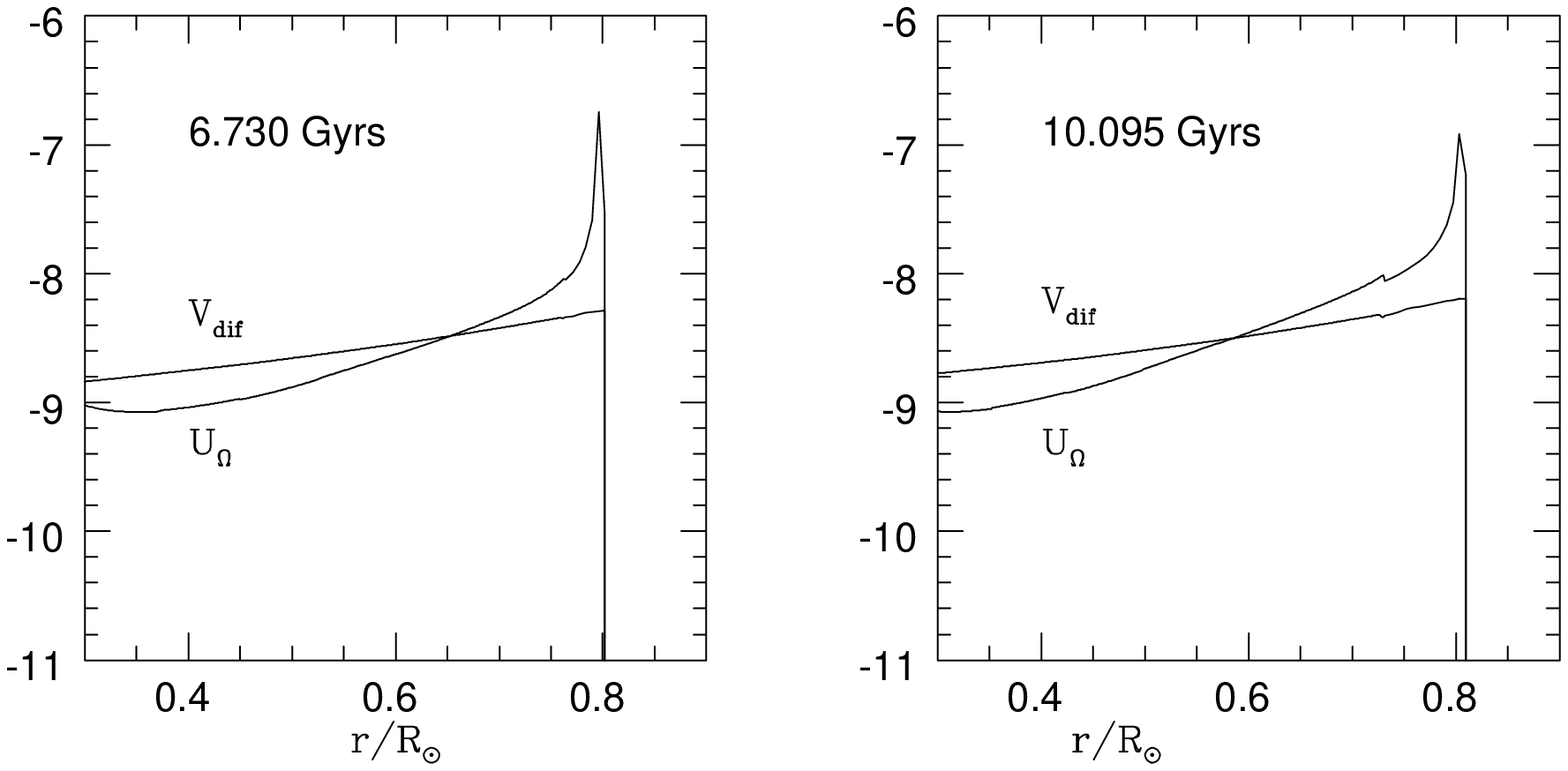}
\centerline{\textbf{Figure 2. b.} Same figure as Figure 2.a. for a $0.70 M_{\odot}$ star.}
\end{figure*}

\begin{figure*}[p]
\centering
\includegraphics[width=13.4cm]{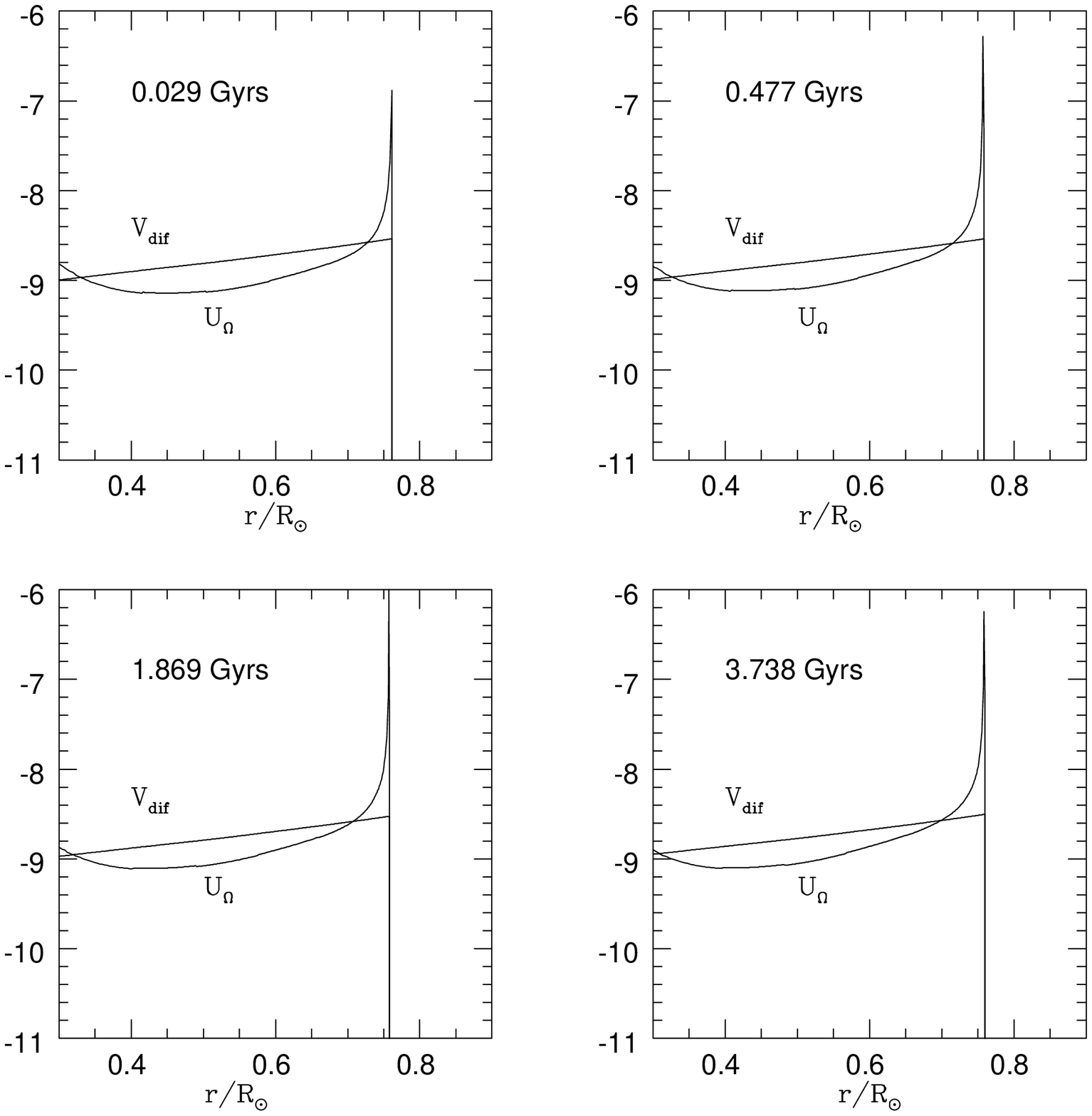}
\includegraphics[width=13.4cm]{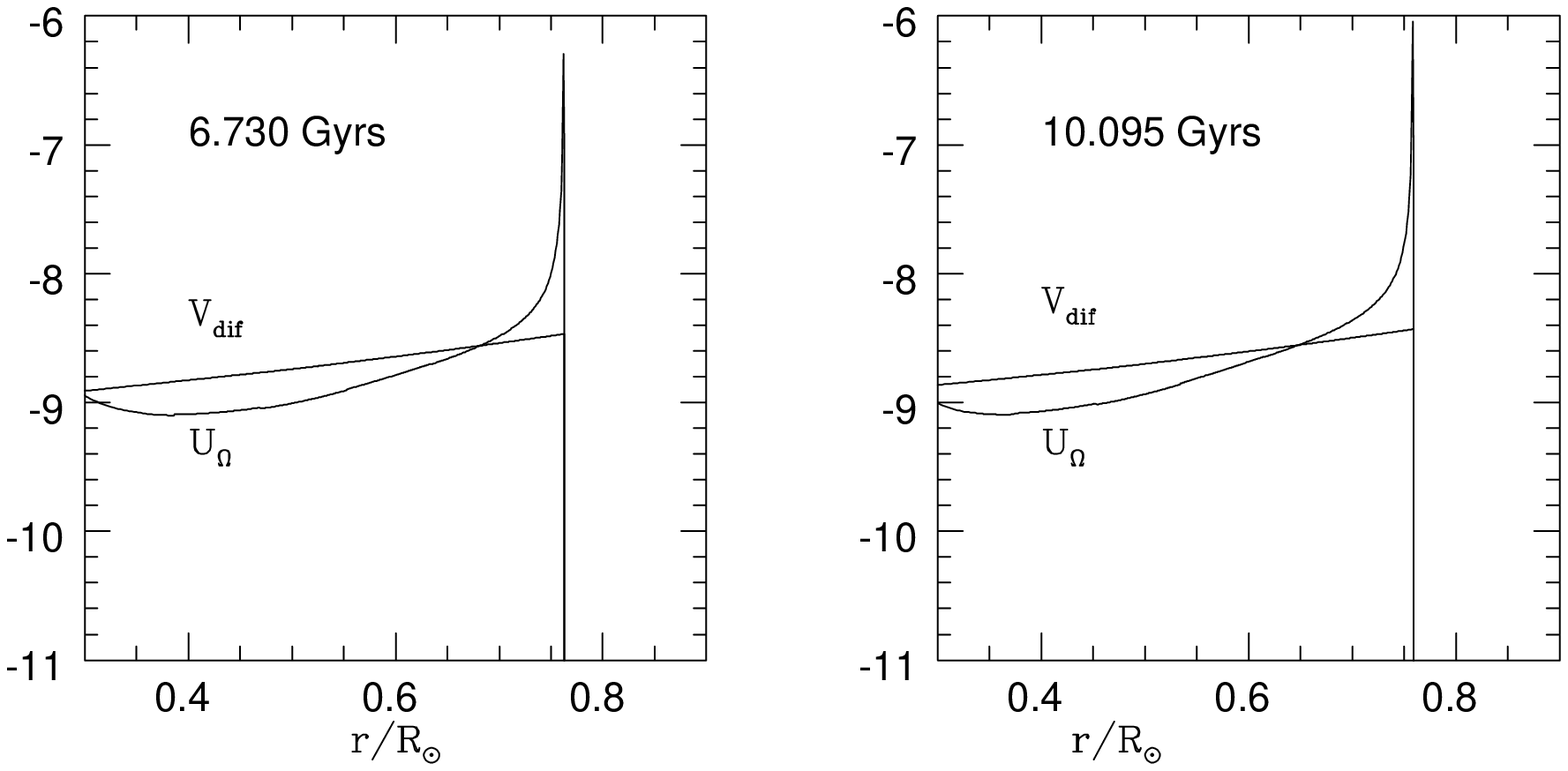}
\centerline{\textbf{Figure 2. c.} Same figure as Figure 2.a. for a $0.65 M_{\odot}$ star.}
\end{figure*}

Figure 3 presents the lithium abundance variations with time for a bunch 
of different rotational velocities in a $0.70 M_{\odot}$ star. For small
 rotation velocities, the lithium abundance variations are mainly due 
to microscopic diffusion, decreasing for increasing rotation velocity.
For velocities larger than 6 km.s$^{-1}$ the depletion is mainly due 
to nuclear destruction, which now increases with increasing velocity.
For velocities between 2.5 and 7 km.s$^{-1}$, the resulting dispersion 
is of order 0.06 dex, going up to .12 dex if the 7.5 km.s$^{-1}$ 
possibility is included.

\begin{figure*}[p]
\centering
\includegraphics[width=10cm]{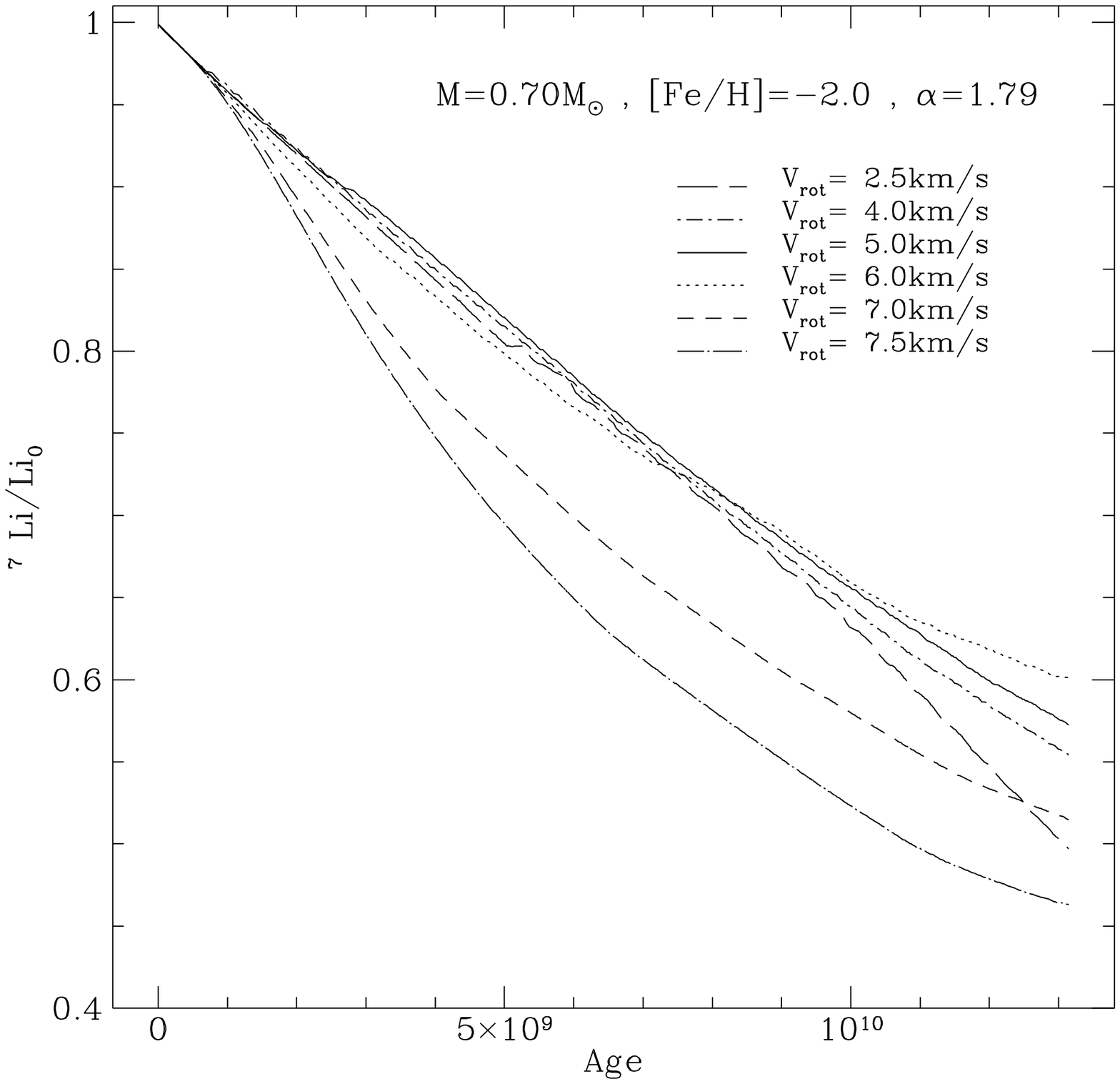}
\caption{Lithium abundance variations at the surface of
 a $0.70 M_{\odot}$ star for rotation velocities between 2.5 and 
7.5 km.s$^{-1}$. The minimum depletion is obtained for 6 km.s$^{-1}$.
For smaller velocities, depletion is mainly due to microscopic 
diffusion while for larger velocities it is mainly due to nuclear
destruction. Due to the self-regulating process, the obtained dispersion
is quite small.}
\end{figure*}

Figure 4 (a, b, c) presents the lithium abundance variations with
time in the outer layers of the four low-metallicity stars 
([Fe/H] = -2) with three different rotation velocities: 2.5, 5.0 and
7.5 km.s$^{-1}$. 
The $0.65 M_{\odot}$ and $0.70 M_{\odot}$ stars
show very similar abundances and their variations with 
the rotation velocity is small except for the largest
value of 7.5 km.s$^{-1}$.
In the $0.75 M_{\odot}$ star
the lithium abundance is more sensitive to the rotation velocity
due to the smaller diffusion time scale. For small velocities the
lithium value in the $0.75 M_{\odot}$ star is smaller than in
cooler stars while it is larger for velocities larger than
5 km.s$^{-1}$. For the cooler $0.60 M_{\odot}$ star, lithium is always more 
depleted due to the small distance between the bottom
of the convective zone and the nuclear destruction layer.

\begin{figure*}[p]
\centering
\includegraphics[width=13.4cm]{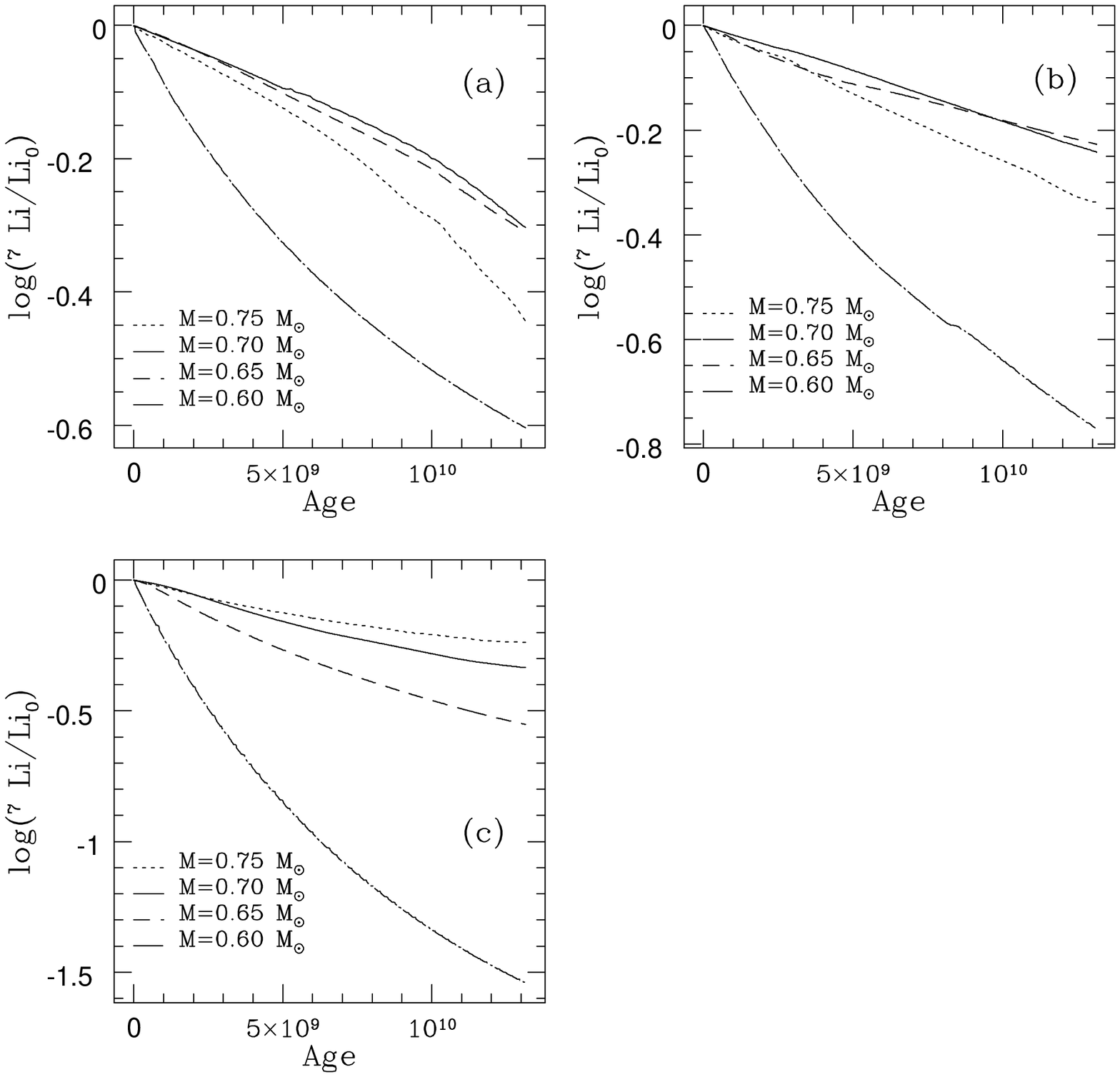}
\caption{Lithium abundance variations at the surface of
 the four halo stars, with three different
rotation velocities : (a) : 2.5 km.s$^{-1}$, (b) : 5 km.s$^{-1}$, (c)
: 7.5 km.s$^{-1}$.}
\end{figure*}
 
Figure 5 (a, b, c, d) displays, as an example, the abundance 
profiles for lithium
inside the four halo stars, for a rotation velocity of 6 km.s$^{-1}$.

\begin{figure*}[p]
\centering
\includegraphics[width=13.4cm]{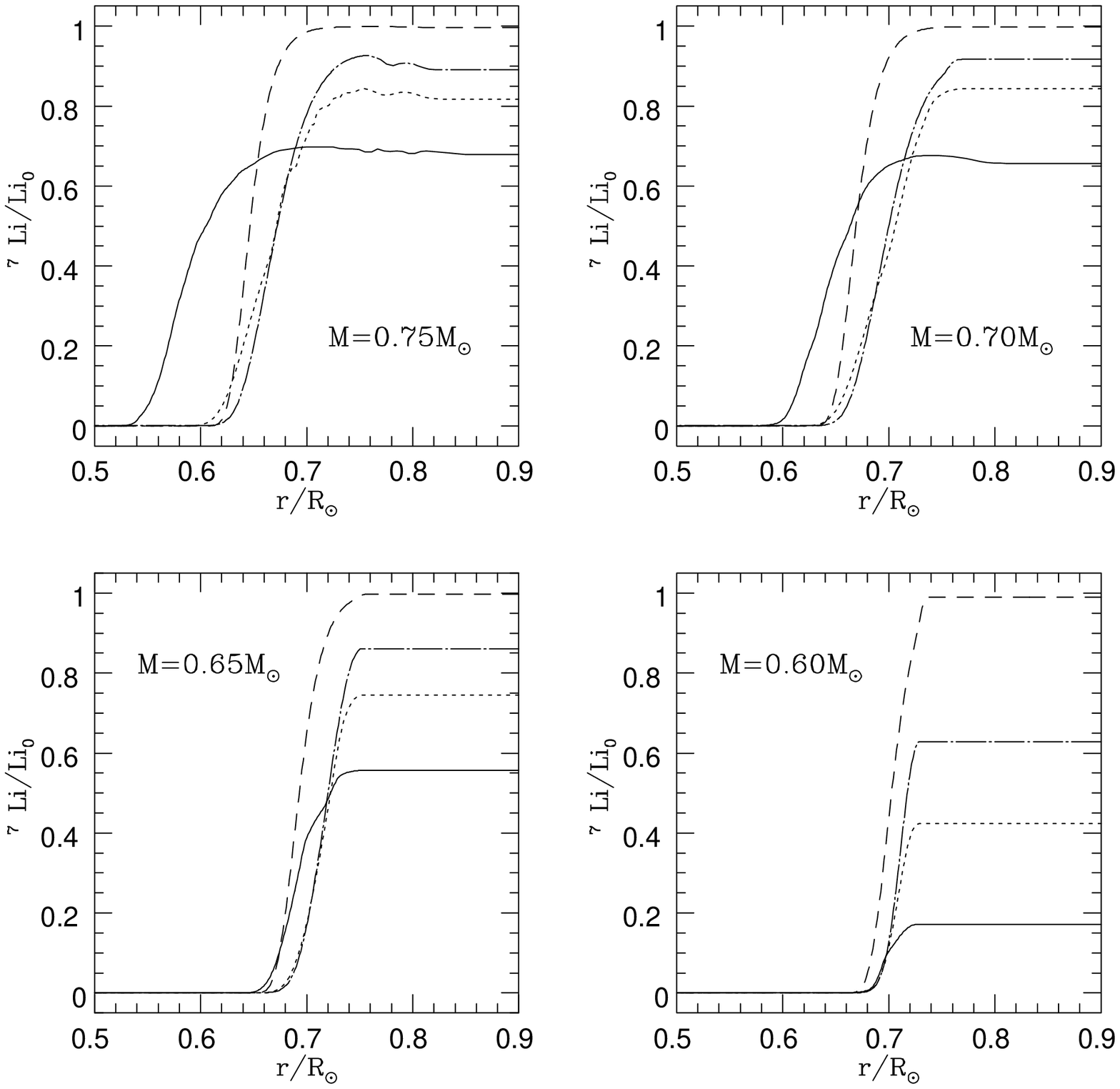}
\caption{Lithium profiles inside the
 four halo stars, with a rotation velocity of
 6 km.s$^{-1}$. Dashed line : 0.029 Gyrs, long dashed dotted line :
 1.869 Gyrs, dotted line : 3.738 Gyrs and solid line : 10.095 Gyrs.}
\end{figure*}

\subsection{Influence of metallicity and mixing length parameter
on the lithium abundances}

We have tested the influence of metallicity on the lithium abundance
variations for all the stars, with two
different metallicities $[Fe/H]=-2.0$ and $[Fe/H]=-3.5$. The curves
representing the lithium variations with time are plotted in Figure 6.
These curves are
very close, which leads to the conclusion of a very small dependence of
the lithium abundance on metallicity. 

\begin{figure*}[p]
\centering
\includegraphics[width=9.9cm]{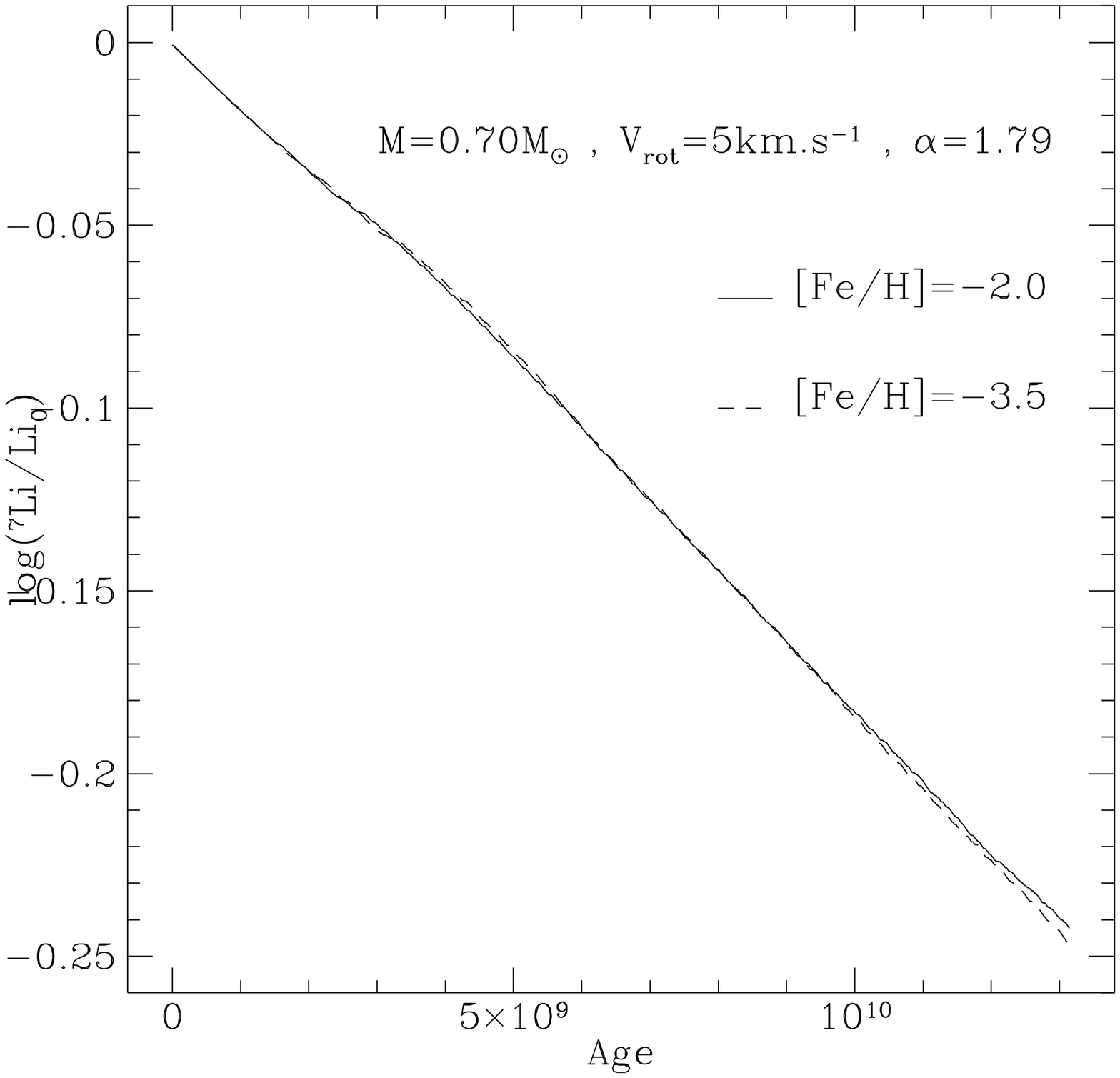}
\caption{Lithium abundance variation with time for $0.70 M_{\odot}$
  halo stars with two different values of the metallicity.}
\end{figure*}

We have also tested the influence 
of varying the mixing length parameter used for
the computation of the convective zone: the results are
presented in Figure 7. The differences are very small : the
results are not sensitive to small variations of the depth of the
convective zone.

\begin{figure*}[p]
\centering
\includegraphics[width=9.9cm]{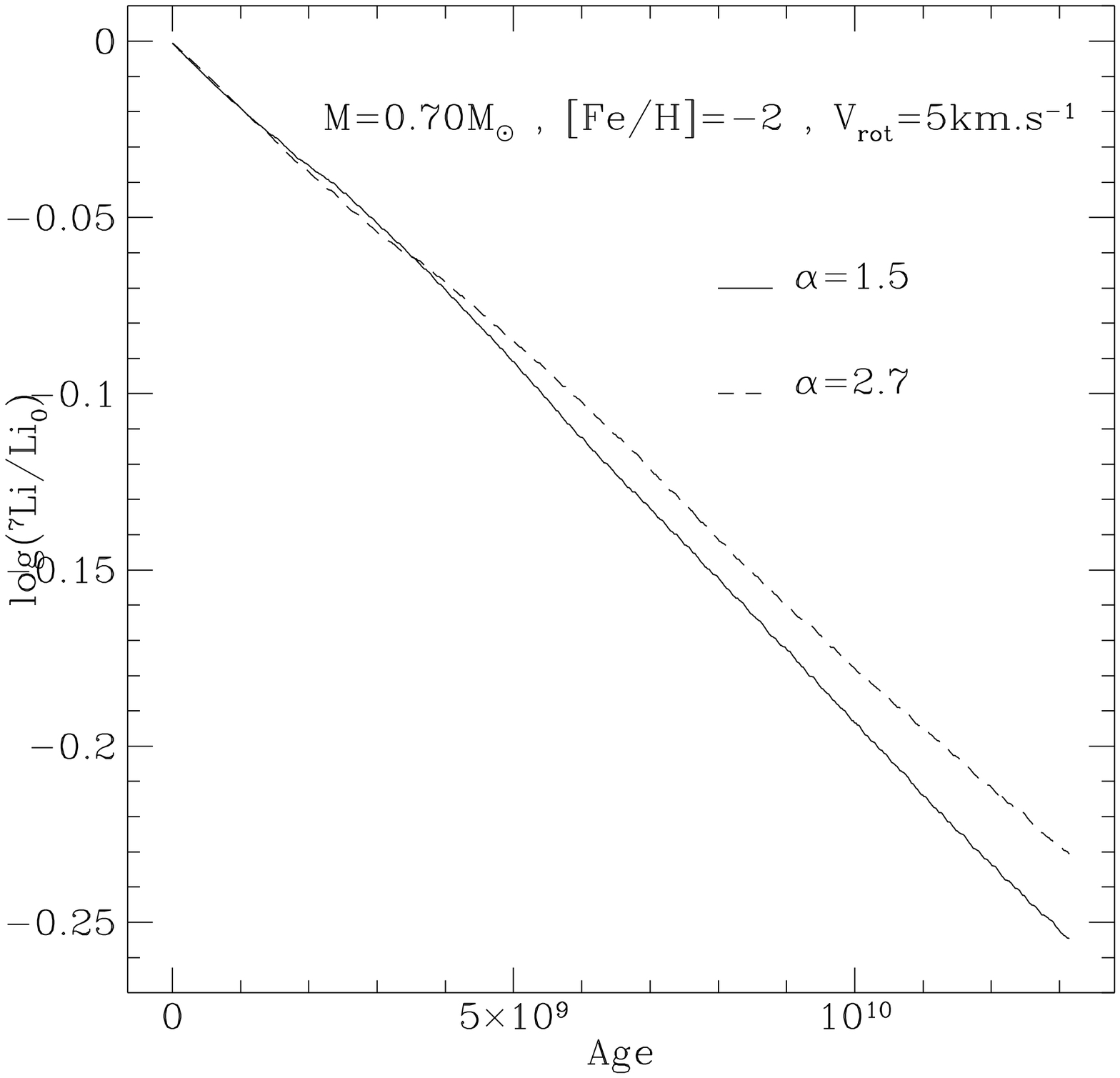}
\caption{Lithium abundance variation with time for $0.70 M_{\odot}$
  halo stars with
two different values of the mixing length parameter.}
\end{figure*}

\subsection{The case of cool stars : introduction of a tachocline}
 
The lithium depletion obtained for the $0.60 M_{\odot}$ star, 
which is
cooler than the lithium plateau, is not sufficient 
to account for the observations,
except for a rotation velocity of 7.5 km.s$^{-1}$.
The observations show, for these effective temperatures, a lithium
depletion by at least a factor of 10, while the computations
obtained by using the same physical inputs as those introduced for the
three
others stars give a lithium depletion of only 60\% for 
2.5 km.s$^{-1}$ and 80\% for 5 km.s$^{-1}$. 

However, at such effective
temperatures, a tachocline layer is supposed to take place below the
convective zone, as in the Sun (Spiegel \& Zahn \cite{Spiegel92}). 
We introduced in the computations an exponential
mixing just below the convective region to reproduce the effects of
this tachocline. The width of the mixed layer was taken as 0.01 R$_{*}$ and
the diffusion coefficient was 100 cm$^2$.s$^{-1}$ below the convective zone, 
decreasing exponentially underneath.

We then obtained the lithium variation with time as plotted on
Figure 8 for a velocity of 5 km.s$^{-1}$, giving a lithium depletion
by a factor 20, consistent with the observations.
We have further checked that such a narrow tachocline layer, if present
in hotter stars, would not modify the previously obtained results.

\begin{figure*}[p]
\centering
\includegraphics[width=9.9cm]{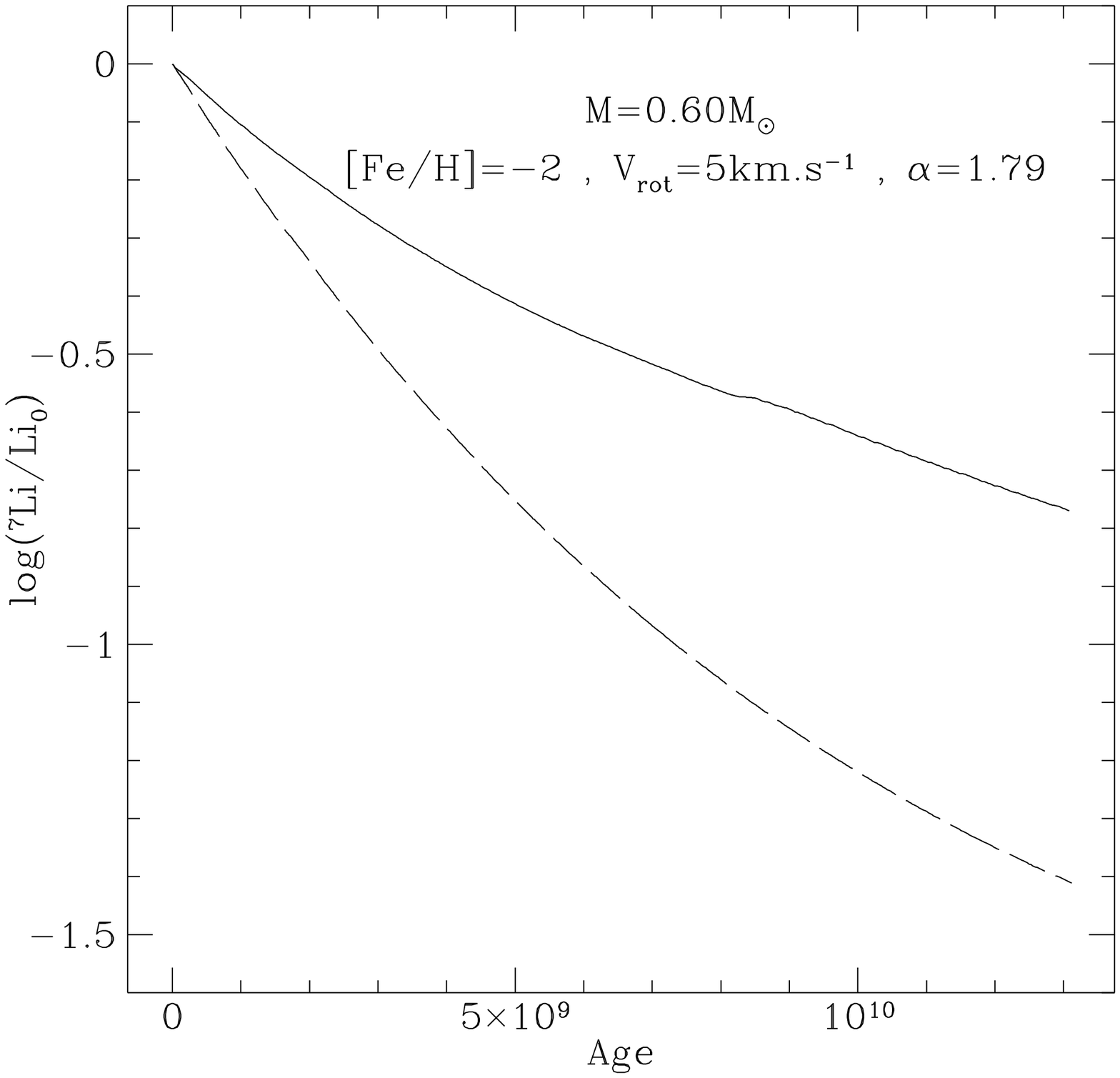}
\caption{Lithium abundance variation with time for $0.60M_{\odot}$
  halo stars with and without tachocline. Solid line :
  stellar model
  without tachocline, dashed line : stellar model with tachocline.}
\end{figure*}

\subsection{Abundance variations of $^6$Li}

We have done the same computations for $^6$Li as for $^7$Li in the 
plateau stars. Figure 9 displays the results for a $0.70 M_{\odot}$
star with metallicity [Fe/H] = -2 and the same values of the 
rotation velocity
as for $^7$Li in Figure 3. The results differ for the two isotopes
because $^6$Li is more easily destroyed by nuclear reactions than $^7$Li, while
the influence of microscopic diffusion is about the same.
For small rotation velocities, less than 7 km.s$^{-1}$, $^6$Li is
destroyed by about a factor of two. For larger velocities it goes
up to a factor of 5. In any case, if $^6$Li is present at the beginning
of the stellar evolution, a non negligible amount remains at the age
of halo stars.

\begin{figure*}[p]
\centering
\includegraphics[width=10.cm]{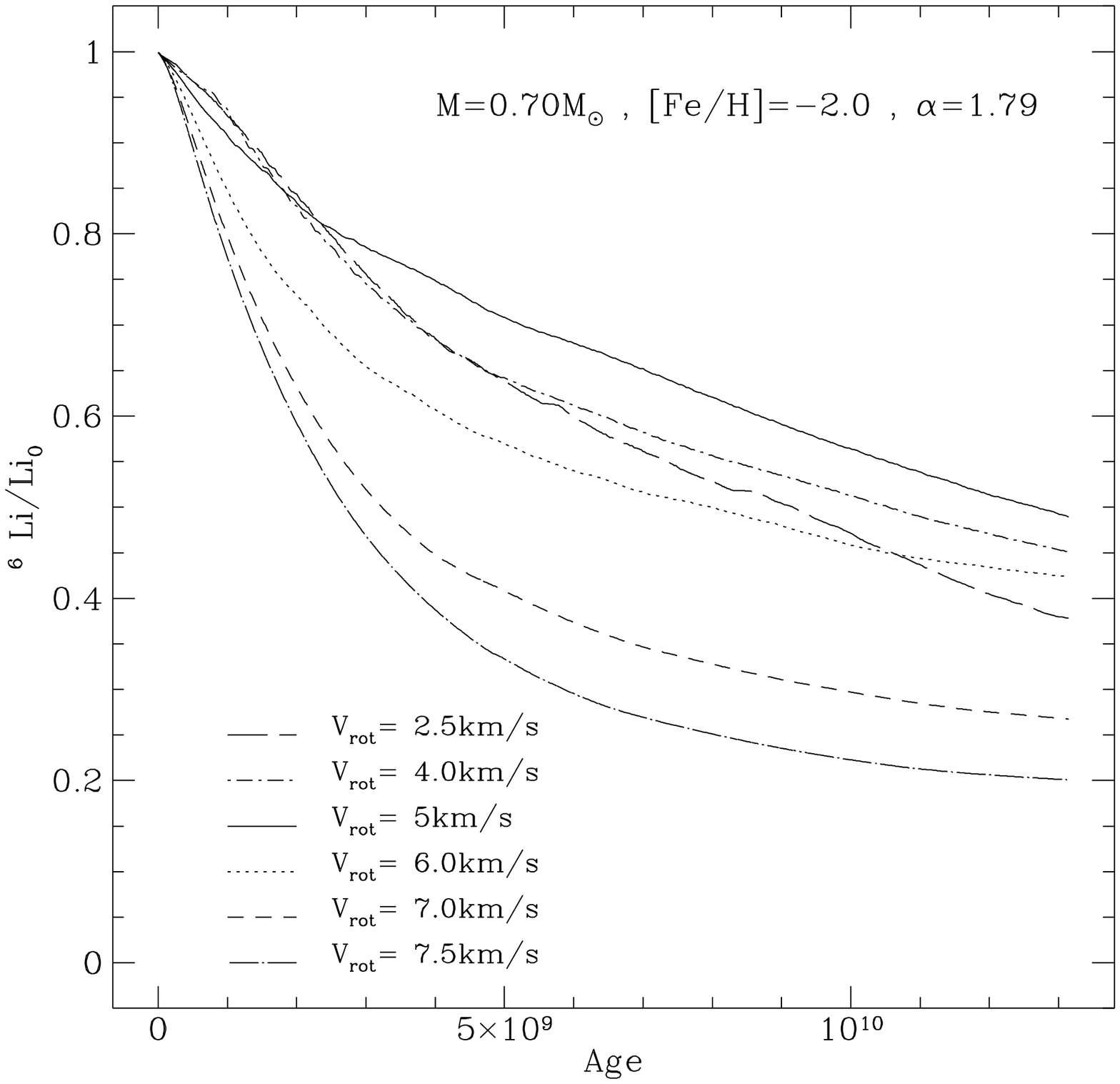}
\caption{Lithium 6 abundance variations at the surface of
 a $0.70 M_{\odot}$ star for rotation velocities between 2.5 and 
7.5 km.s$^{-1}$.}
\end{figure*}

\section{Discussion}

\begin{table*}
\begin{center}
\begin{tabular}{|c|c|c|c|c|c|c|c|c|c|}
\hline
\multicolumn{2}{|c|}{Masse}
& \multicolumn{2}{c|}{M=0.75 M$_{\odot}$}
& \multicolumn{2}{c|}{M=0.70 M$_{\odot}$}
& \multicolumn{2}{c|}{M=0.65 M$_{\odot}$}
& \multicolumn{2}{c|}{M=0.60 M$_{\odot}$} \\
\cline{3-10}
\multicolumn{2}{|c|}{}& $\displaystyle \log \frac{Li}{Li_0}$ &
$\displaystyle T_{{\rm eff}}$ &$\displaystyle \log \frac{Li}{Li_0}$ &
$\displaystyle T_{{\rm eff}}$ &$\displaystyle \log \frac{Li}{Li_0}$ &
$\displaystyle T_{{\rm eff}}$ & $\displaystyle
\log \frac{Li}{Li_0}$ & $\displaystyle T_{{\rm eff}}$ \\
\hline
V=2.5 km.s$^{-1}$
& 10 Gyrs &-0.29 &6393 &-0.20 &6059 &-0.22 &5718 &-0.52 &5383 \\
\cline{2-10}
 & 12 Gyrs &-0.38 &6524 &-0.26 &6139 &-0.28 &5773 &-0.57 &5421 \\
\hline
V=5.0 km.s$^{-1}$
& 10 Gyrs &-0.26 &6391 &-0.18 &6059 &-0.18 &5715 &-0.64 &5185 \\
\cline{2-10}
  & 12 Gyrs &-0.31 &6473 &-0.22 &6133 &-0.21 &5770 &-0.73 &5226 \\
\hline
V=7.5 km.s$^{-1}$
& 10 Gyrs &-0.21 &6397 &-0.28 &6059 &-0.46 &5720 &-1.34 & 5385 \\
\cline{2-10}
  & 12 Gyrs &-0.23 &6507 &-0.32 &6136 &-0.52 &5774 &-1.47 &5423 \\
\hline
\end{tabular}
\caption{Lithium depletion and effective temperatures of stars for the
  four considered masses, three different rotation velocities and two
  ages. Here no tachocline is included in the models.}
\end{center}
\end{table*}

The results obtained in this paper are compared to
the observational points and error bars given by 
Bonifacio \& Molaro (\cite{Bonifacio97}) in
Figure 10 (a) and (b). 
The solid curve drawn in Figure 10(a) represents
averaged depletion values for rotation velocities
between 2.5 and 7.5 km.s$^{-1}$. The original lithium
value is A(Li) = 2.5.
In figure 10 (b), two extreme curves are given, again
with the assumption of rotation velocities between
2.5 and 7.5 km.s$^{-1}$. The expected dispersion
is very small (of order 0.1 dex) at the middle
of the lithium plateau. It becomes larger at both extremes:
for the hotter side, it is due to the shorter diffusion
time scale and for the cooler side to the
proximity of the nuclear destruction layers.

A complete treatment of the 
self-regulating process described in
section 4-3 would need numerical simulations
which are in preparation. Here we have
assumed that this effect significantly reduces
both the circulation and the diffusion as the
circulation time scale is smaller than the diffusion
time scale while it is supposed to proceed unaltered
below. Such a simple scheme gives results consistent with the
lithium observations in halo stars. No mass loss has been
taken into account, which means that the mass loss rate
is assumed smaller than $10^{-13} M_{\odot}.yr^{-1}$ 
(Vauclair \& Charbonnel \cite{Vauclair95}).

Two important assumptions have been used in these 
computations : the stellar rotation velocity was
taken as a constant during evolution and the
differential rotation inside the stars was assumed
negligible.

The first assumption is impossible to check directly, as we
have no access to the past rotational history of 
halo stars. It is generally assumed that they suffer
the same rotational breaking as pop I stars. However
they have not been formed in the same way, not being
in the same galactic sites, and their kinematics is different. There is
no reason why halo stars should behave exactly in the same way as open
cluster stars. If the rotational velocity of halo stars was larger in
the past, we would expect a larger lithium destruction before the
self-regulating process we have described takes place. This cannot be
excluded, so that the result we give below has to be taken as the lower
limit of the primordial value.

The second assumption is supported by recent helioseismic results,
which show that the internal
rotation of the Sun follows a solid body law except in the
convective zone. We may suppose that the same occurs in halo
stars. The fact that the obtained results nicely 
reproduce the observations argue in favour of this 
hypothesis. Note that the most efficient way to reach this situation seems
to be the presence of a very small magnetic field (Mestel et
al. \cite{Mestel87},
Charbonneau \& MacGregor \cite{Charbonneau92}) which, as discussed by
Mestel (\cite{Mestel99}),
would suppress
differential rotation without altering large-scale motion. Such a
physics has to be further studied. 

It is possible, however, that in some specific stars, due to
gravitational interaction with other stars or to some unknown
other process, differential rotation does occur. This could 
explain the existence of some stars with no observed lithium
below the plateau : lithium would have been destroyed there
by nuclear reactions due to extra mixing. It could also explain 
why some dispersion
is observed in globular cluster stars while it is not
observed in field stars (Boesgaard \& Deliyannis
\cite{Boesgaard98}). 

Futher computations have to be done to describe more precisely
the hydrodynamical situation which occurs when the 
$\displaystyle \mu $-currents oppose the 
$\displaystyle \Omega $-currents below the convective zones
of slowly rotating stars. Applications to lithium in open clusters are
underway. 
The results presented here are
very encouraging in this respect. They show that the plateau
can be reproduced with no ``ad hoc" adjustment of any 
parameter, with an overall lithium depletion slightly
smaller than a factor two. In this case $^6$Li is still
present in the plateau stars. The primordial lithium
value is then found as 2.5 $\pm$ 0.1, consistent with
the recent determinations of D/H and $^4$He/H (see Burles \& Tytler
\cite{Burles98}). These results give a baryonic number $\eta = (5 \pm
1).10^{-10}$ and a baryonic density $\Omega_b h^2$ between 0.015 and
0.022.

\begin{figure*}[p] 
\centering
\includegraphics[width=10cm]{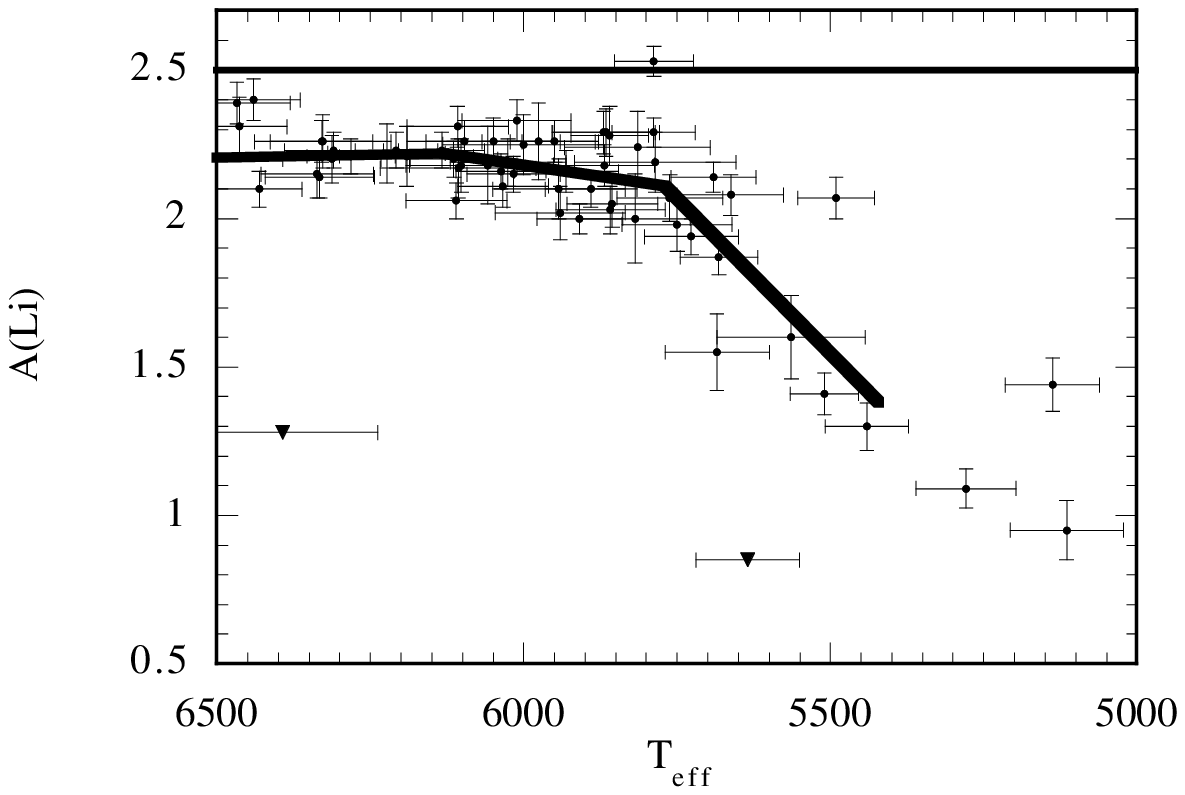}
\includegraphics[width=10cm]{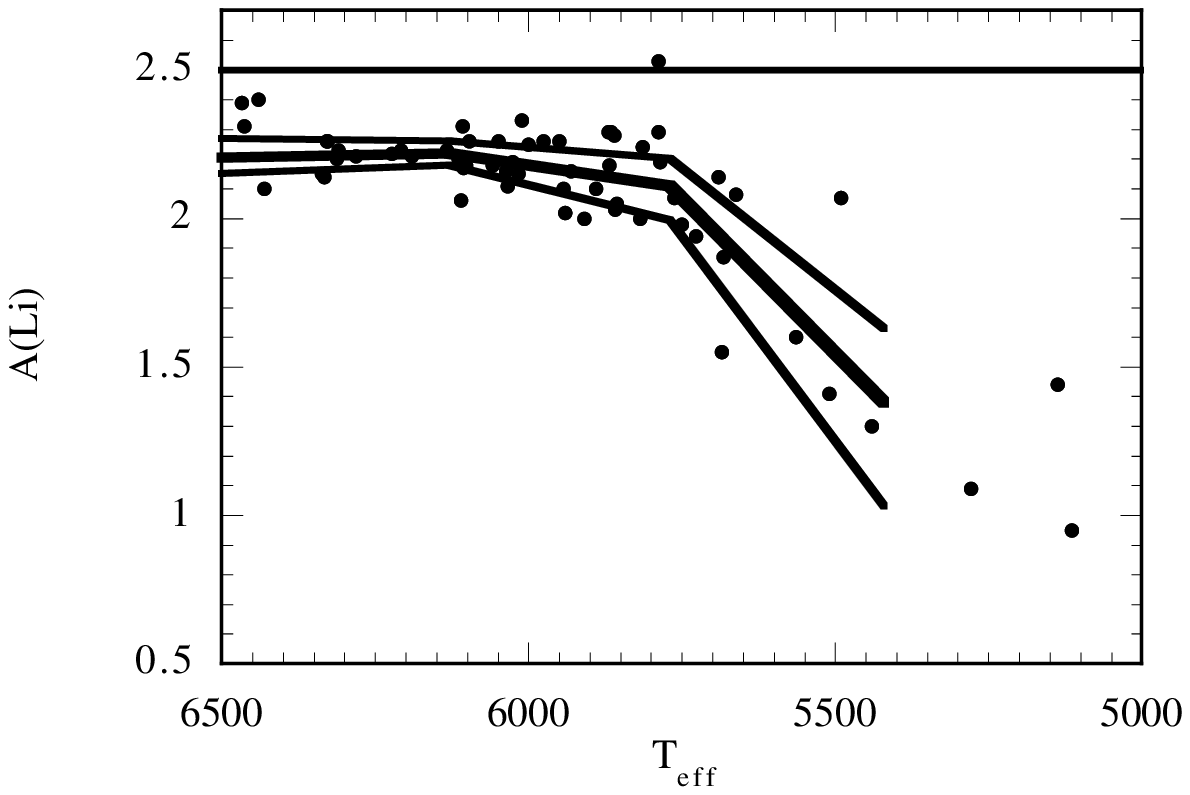}
\caption{Comparison between the computations of lithium depletion and
  the observational values of Bonifacio \& Molaro
  (\cite{Bonifacio97}). (a) : average curve obtained for rotation
  velocities between 2.5 and 7.5 km.s$^{-1}$. Here a tachocline is
  introduced as explained in section 5.4. The original lithium value
  is choosen as 2.5 to fit the observations. (b) : expected dispersion
  if the rotation velocities lie between 2.5 and 7.5 km.s$^{-1}$. The
  observational error bars are not shown for clarity. }
\end{figure*}

\end{document}